%% file: main.tex
 \documentclass[final, 3p, times, fleqn]{elsarticle}

\usepackage{newtxtext}  
\usepackage{lmodern} 

\usepackage{amssymb}
\usepackage{lipsum}
\usepackage[utf8]{inputenc} 
\usepackage[T1]{fontenc}    
\usepackage{url}            
\usepackage{booktabs}       
\usepackage{amsfonts}       
\usepackage{nicefrac}       
\usepackage{microtype}      
\usepackage{xcolor}         
\usepackage{amsmath}
\usepackage{bm}
\newcommand\norm[1]{\left\lVert#1\right\rVert}

\usepackage{amssymb}
\usepackage{mathtools}
\usepackage{amsthm}
\usepackage{multirow}
\usepackage{graphicx}
\usepackage[hidelinks]{hyperref} 
\usepackage{calc}
\usepackage{cancel}
\usepackage{enumitem}

\journal{Computer Methods in Applied Mechanics and Engineering}

\begin{document}

\begin{frontmatter}



\title{A Physics-Augmented GraphGPS Framework for the Reconstruction of 3D Riemann Problems from Sparse Data}


\author[first]{Rami Cassia}
\author[first]{Rich Kerswell}
\affiliation[first]{organization={Department of Applied Mathematics and Theoretical Physics, University of Cambridge},
            addressline={Wilberforce Rd}, 
            city={Cambridge},
            postcode={CB3 0WA}, 
            country={UK}}

\begin{abstract}
In compressible fluid flow, reconstructing shocks, discontinuities, rarefactions, and their interactions from sparse measurements is an important inverse problem with practical applications. Moreover, physics-informed machine learning has recently become an increasingly popular approach for performing reconstructions tasks. In this work we explore a machine learning recipe, known as GraphGPS, for reconstructing canonical compressible flows known as 3D Riemann problems from sparse observations, in a physics-informed manner. The GraphGPS framework combines the benefits of positional encodings, local message-passing of graphs, and global contextual awareness, and we explore the latter two components through an ablation study. Furthermore, we modify the aggregation step of message-passing such that it is aware of shocks and discontinuities, resulting in sharper reconstructions of these features. Additionally, we modify message-passing such that information flows strictly from known nodes only, which results in computational savings, better training convergence, and no degradation of reconstruction accuracy. We also show that the GraphGPS framework outperforms numerous machine learning benchmarks. 
\end{abstract}




\end{frontmatter}

\section*{Highlights}
\begin{itemize}[leftmargin=0.35cm]
    \item Investigation of GraphGPS for flow reconstruction.
    \item Reconstruction of 3D Riemann problems.
    \item Shock-aware message-passing module.
    \item Modified message-passing mechanism that is more interpretable, computationally efficient, and stable.
    \item Reconstruction is agnostic to initialization of unknown points.
\end{itemize}




\input{Introduction.tex}

\input{Priorwork.tex}
\input{Methodology.tex}
\input{Experiments.tex}
\input{Conclusion.tex}

\section*{Code and Data}

Code and data can be accessed via the
\href{https://github.com/RamiCassia/GraphGPS-for-Sparse-Reconstruction}{{\textcolor{blue}{following link}}}.

\section*{Acknowledgments}

Rami Cassia acknowledges the financial support provided by the Cambridge Commonwealth, European and International Trust.

\clearpage
\bibliography{bibli}


\newpage
\appendix


\input{Appendix_A}

\newpage
\input{Appendix_B}
\newpage
\input{Appendix_C}

\clearpage
\input{Appendix_D}







\end{document}

%% file: Introduction.tex
\section{Introduction}

A common inverse problem is the reconstruction of a set of variables from sparse data. Applications of this inverse problem span a range of fields such as imaging, audio and speech processing, medicine, structural engineering, seismology, finance, astronomy, and fluid dynamics. Particular applications in fluid dynamics include climate modeling, oceanography, and aerospace engineering. In aerospace engineering, for example, it is common practice to mount pressure sensors on a wing during wind-tunnel testing in order to reconstruct the surrounding flow-field. 

In this paper we are interested in the sparse reconstruction of the 3D Riemann problems, which are a set of problems that investigate the interaction of elementary waves (shocks, rarefactions, contact discontinuities) as governed by the compressible Euler equations. We are specifically interested in performing the reconstruction using a physics-informed machine learning (ML) model that is a) graph-based and b) contextually-aware. A model that is graph-based can leverage data directionality, which we hypothesize is beneficial to reconstruction where information should flow from known to unknown points. A model that is contextually-aware is capable of capturing long-range dependencies in the data, which is crucial in flows where there are long-range correlations due to the presence of shocks, contact discontinuities and rarefactions. As such we briefly review graph-based and contextually-aware models.

Graph-based models are those that operate on graphs comprising of vertices (or nodes) and edges. Graph models typically involve message-passing, which refers to the nodal aggregation of messages or information from neighbours. A key property of message-passing is permutation-invariance, meaning features can be updated in any node ordering. Furthermore, message-passing inherently captures local dependencies in data. Graph neural networks (GNNs) and message-passing were first introduced by \citet{scarselli2008graph}. Graph convolutional networks were then introduced by \citet{kipf2016semi}, which perform the message-passing aggregation through convolutional operations, increasing efficiency and interpretability. \citet{velivckovic2017graph} improve over GCNs by introducing graph attention networks (GATs) which use attention scores in the aggregation process. The attention mechanism reduces oversmoothing typically encountered in GCNs. \citet{hamilton2017inductive} also improve over GCNs by using sampling-based aggregation (graphSAGE), enabling scalability to larger graphs. \citet{xu2018powerful} use a MLP-based aggregation of neighbours in their graph isomorphism network (GIN). They show that their network is as expressive as the Weisfeiler-Lehman test (a graph isomorphism test), which enables their model to better distinguish graph structures.

Contextually-aware models are those that capture long-range dependencies, typically through global attention mechanisms. The origins of attention mechanisms can be traced to the work of \citet{bahdanau2014neural} and \citet{luong2015effective} who introduce additive and dot-product attention in their recurrent-based architectures for neural machine translation. However, the major breakthrough came when the transformer architecture, presented by \citet{vaswani2017attention}, completely removed recurrence through usage of sinusoidal positional encodings and multi-head self attention. This novel architecture is parallelizable over sequences which significantly reduces training time. They generalize the attention mechanism in terms of querys, keys and values where the multiple heads allows learning of different aspects of the input. Their attention is a scaled dot-product attention where the scaling prevents exploding gradients. The transformer breakthrough led to numerous milestones in language modeling and vision \cite{radford2018improving, kenton2019bert, alexey2020image}. There has also been work on reducing the computational complexity of transformers from quadratic to linear. \citet{katharopoulos2020transformers} achieve linear complexity by expressing self-attention as a linear dot-product of kernel feature maps and by making use of the associativity of matrix multiplication. \citet{choromanski2020rethinking} also achieve linear complexity by approximating kernelized attention using random feature maps. Recently, there has been work on replacing attention with state space models (SSMs) that are contextually-aware to achieve linear complexity. \citet{gu2023mamba} do this by making the SSM parameters input-dependent, allowing their model to selectively propagate or forget information along a sequence depending on the current token. Their model, named Mamba, is implemented using a hardware-aware algorithm to address the computational inefficiencies caused by the recurrent-nature of the model. \citet{dao2024transformers} later develop a theoretical framework that connects SSMs with certain variants of attention - enabling the design of a newer model, Mamba-2, that also leverages the modern hardware optimizations of attention. There has also been work on extending transformers and selective-SSMs to handle graphical data \cite{dwivedi2020generalization, kreuzer2021rethinking, ying2021transformers, behrouz2024graph, wang2024graph}. 

In this paper, we are interested in using the framework proposed by \citet{rampavsek2022recipe} in performing our fluid reconstruction task. The framework, known as GraphGPS, comprises three main components: a local message-passing mechanism via graphs, a global attention mechanism, and positional/structural encodings. The contributions of our paper are as follows:

\begin{itemize}[leftmargin=0.35cm]

    \item To the best of our knowledge, this is the first study on the use of GraphGPS for flow field reconstruction. In particular, we make use of the modularity of the framework to explore the effect of different global attention mechanisms and different message-passing mechanisms. We also show superior performance of GraphGPS against other ML architectures.
    
    \item For the message-passing module, we also introduce a modified GAT layer, known as SA-GATConv, that leverages information on shocks and discontinuities in computing the local attention weights. We show that this produces sharper reconstructions in regions of shocks and discontinuities. 

    \item We further leverage the graphical structure during message-passing such that a node that is initially unknown cannot propagate information to neighbours unless information has first been propagated to it from a known node. Compared to a naive form of message passing where all nodes (known or unknown) propagate information to each other, this form of \textit{guided message-passing} is more interpretable, reduces computational overhead, and improves convergence.

    \item We ensure that in the projected space of GraphGPS, the representation of unknown nodes is zeroed such that the only information available is through their positional encoding. We achieve this through a masked projection of the input, where the masking is based on whether a node is known or unknown. This also ensures the reconstruction is agnostic to the initialization values at unknown nodes. We show empirically that this setup performs just as well as an unmasked (normal) projection of the input where feature propagation \cite{rossi2022unreasonable} has been used to provide an initial guess at unknown nodes. 

    \item To the best of our knowledge, this is the first study on the 3D Riemann problems in the SciML literature.
    
\end{itemize}


This paper is organized as follows. In Section 2, we review prior work on flow field reconstruction from sparse measurements. In Section 3, we mathematically formulate our reconstruction problem, and outline the GraphGPS architecture and its various aspects. In Section 4, we show our experimental findings. Finally, we finish with conclusions in Section 5.



%% file: Priorwork.tex
\section{Prior Work on Flow Reconstruction}

One popular approach for reconstruction in fluid dynamics is through proper orthogonal decomposition (POD), which involves finding orthonormal basis vectors that minimize the difference between the known data points and their projection onto a lower dimensional subspace. Gappy POD uses a mask to enforce the POD to be applied on known points only, and the basis vectors can then be used to estimate the variables at all grid points.  The method was first introduced by \citet{everson1995karhunen} for reconstructing human face images, and later used by \citet{bui2004aerodynamic} in an aerodynamic setting where the pressure field around aerofoils were reconstructed from sensors on the aerofoil surface. \citet{vendl2010proper} also used the method to recover the transonic flow over a wing-body configuration.  The downside of POD is that it assumes a linear decomposition and may struggle to capture non-linear flow behaviour. Other classical methods are interpolation-based. One such method constructs a function that is a weighted sum of radial basis functions (RBFs). Another such method is Kriging interpolation, which is a probabilistic method that estimates values at unmeasured locations based on spatial correlations in the data, whilst also providing uncertainty estimates. \citet{huang2024surrogate} investigate the use of Kriging and RBFs in the reconstruction of hypersonic flow through a nozzle. 


More recently machine learning has become a powerful paradigm for performing reconstruction. \citet{duthe2023graph} use a graph neural network to simultaneously reconstruct both the pressure and velocity fields around aerofoils from the surface pressure distributions. Their model is also capable of predicting global properties of the flow such as inflow velocity and angle of attack. They make use of feature propagation to initialize unknown nodes. \citet{xu2023practical} use a physics-informed neural network to reconstruct the wake flow past a cylinder. They place emphasis on the importance of training procedures in drastically reducing the number of training epochs required, by using a cosine-annealing learning rate schedule with warm restarts. \citet{kim2024gappy} use a gappy autoencoder to perform reconstruction via non-linear manifold representations. The method shows improvement over gappy POD (which utilizes linear subspaces), as tested on diffusion, advection and wave problems. They also assess the impact of different sensor placements on reconstruction accuracy. \citet{peng2024rapid} use a physics-informed graph neural network to reconstruct high speed transient flows. They use a fully-connected neural network to map the sparse data to the entire domain, and then assign the mapped data to graph nodes to facilitate information exchange. Finally the output state variables are used to compute a PINN loss summed together with initial condition and boundary condition losses. Their model exhibits good spatial and temporal generalization but has low accuracy when reconstructing flow near wave-fronts. \citet{hosseini2024flow} use a physics-informed neural network to reconstruct the wake flow past a cylinder from sparse sensors placed around the domain boundary and cylinder wall. \citet{danciu2024flow} use graph attention convolutional layers to reconstruct the flow in an internal combustion engine where the geometry is time-varying due to piston motion. They use a feature propagation step to leverage information from known nodes to initialize missing features, and use a mask to distinguish between original and propagated data points which allows more effective learning from the sparse inputs. \citet{quattromini2024graph} make use of the adjoint method to compute Reynolds-averaged Navier Stokes (RANS) gradients which are then used as optimization terms during the training of a GNN. They use this hybrid ML-CFD approach to reconstruct the mean flow past cylinders. \citet{nguyen2024flrnet} introduce FLRNet which uses a variational autoencoder to learn a latent representation of the flow-field past a cylinder. The VAE makes use of Fourier feature layers and a perceptual loss term to enhance its ability to learn detailed features of the flow-field. An MLP is then used to correlate sparse sensor measurements to the learned representation. \citet{dang2024flronet} use an operator learning framework to perform reconstruction. The framework utilizes a branch-trunk architecture. The branch network, which uses Fourier neural operators, incorporates sensor measurements at different timestamps. The trunk network uses an MLP to encode the entire temporal domain. Combining the outputs of the branch and trunk nets yields the final reconstruction which is discretization-independent. 


%% file: Methodology.tex
\newcommand*{\Scale}[2][4]{\scalebox{#1}{$#2$}}%
\newcommand*{\Resize}[2]{\resizebox{#1}{!}{$#2$}}%

\section{Methodology}\label{sec:back}

\subsection{Problem Definition}\label{sec:probdef}

We are interested in the following inverse problem:

\textit{Find  $\mathcal{F}_{\Theta}: \left( \mathbf{X} \in \mathbb{R}^{N_{xyz} \times N_{tc}}, \,\,\,\, \mathbf{m} \in \left\{0,1\right\}^{N_{xyz}} \right) \longrightarrow \mathbf{\hat{X}} \in \mathbb{R}^{N_{xyz} \times N_{tc}} \,\,\,\,$ such that $\epsilon\left(\mathbf{\hat{X}}, \mathbf{\bar{X}}\right)$ is minimized, where:
\begin{itemize}[leftmargin=0.35cm]
    \item $\mathbf{X}$, $\mathbf{\hat{X}}$, and $\mathbf{\bar{X}}$ are the sparse, reconstructed, and true flow fields, respectively.
    \item $\mathbf{m}$ is the binary mask which indicates where values are known or unknown.
    \item $\mathcal{F}_{\Theta}$ is the machine learning model with learnable weights  $\Theta$.
    \item $N_{xyz} = N_{x}N_{y}N_{z}$ represents the number of points in 3D space. $N_{tc} = N_{t}N_{c}$ represents the number of features where $N_t$ is number of timesteps and $N_c$ is the number of variables in the physical system.
    \item $\epsilon$ is the error metric between $\mathbf{\hat{X}}$, and $\mathbf{\bar{X}}$ for measuring accuracy (not to be confused with the training loss function).
\end{itemize}}

In this paper,  we are interested in flow fields pertaining to the 3D Euler equations, so $N_c = 5$:
\begin{align}
&\partial_{t}\mathbf{U} + \partial_{x}\mathbf{F}\left(\mathbf{U}\right) + \partial_{y}\mathbf{G}\left(\mathbf{U}\right) + \partial_{z}\mathbf{H}\left(\mathbf{U}\right) = \mathbf{0},\\[6pt]
&{\partial_t}
\begin{pmatrix}
        \rho      \\
        \rho u      \\
        \rho v      \\
        \rho w \\
         E      \\
\end{pmatrix} + {\partial_x}
\begin{pmatrix}
        \rho u      \\
        \rho u^{2} + p \\
        \rho u v \\
        \rho uw \\
        u\left(E + p\right) \\
\end{pmatrix} + {\partial_y}
\begin{pmatrix}
        \rho v      \\
        \rho u v    \\
        \rho v^{2} + p \\
        \rho vw \\
        v\left(E + p\right) \\
    \end{pmatrix} + {\partial_z}
\begin{pmatrix}
        \rho w      \\
        \rho uw    \\
        \rho vw \\
        \rho w^{2} + p \\
        w\left(E + p\right) \\
    \end{pmatrix}
    = \mathbf{0}, \\[6pt]
&E = \frac{1}{2}\rho\left(u^2 + v^2 + w^2\right) + \frac{p}{\left(\gamma - 1\right)},
\end{align}
where $\rho$, $u$, $v$, $w$, $p$, $E$, $\gamma$ are density, $x$-velocity, $y$-velocity, $z$-velocity, pressure, energy and ratio of specific heats, respectively. The boundary conditions are assumed to be Neumann, i.e., $\partial_{\mathbf{n}} \mathbf{U} = \mathbf{0}$. To enforce positivity preservation (i.e., non-negative density and energy, \cite{batten1997choice}) as a hard constraint, we work with the following representation for $\mathbf{X}$ at each spatiotemporal point:
\begin{equation}
    \mathbf{X} = \left( \rho, u, v,w, E - \frac{1}{2}\rho(u^{2} + v^{2} + w^{2}) \right)^T,
\end{equation}
and apply during training, for each spatiotemporal point, the transformation $\mathcal{T}$ after applying $\mathcal{F}_{\Theta}$:
\begin{equation}
    \mathcal{T}: \left( \rho, u, v,w, E - \frac{1}{2}\rho\left(u^{2} + v^{2} + w^{2}\right) \right)^T \longrightarrow  \left( \left|{\rho}\right|, u, v,w, \left|E - \frac{1}{2}\rho\left(u^{2} + v^{2} + w^{2}\right)\right|\right)^T.
\end{equation}
We obtain $\mathbf{\bar{X}}$, the true field, by numerically solving the 3D Euler equations using a 5th-order HLLC-WENO scheme (see \ref{app:hllc} and \ref{app:weno} for more details). The mask $\mathbf{m}$ is randomly generated from a uniform distribution such that 90\% of points are unknown. For element $m_i$ of $\mathbf{m}$:
\begin{equation}
m_{i} = 
\begin{cases}
0 & \text{if} \,\,\,\,\, a_{i}  < 0.9, \\
1 & \text{if} \,\,\,\,\, a_{i} \geq 0.9.
\end{cases}, \,\,\,\,\,\,\,\,\,\,  i \in [1, N_{xyz}], \,\,\,\,\,\,\,\,\,\, a_{i} \sim \mathcal{U}\left(0,1\right).
\end{equation}
The sparse field is then generated from the mask as $\mathbf{X} = \mathbf{m}\odot\mathbf{\bar{X}}$, where $\odot$ is the Hadamard product. For the error metric, we choose:
\begin{align}
\epsilon = \frac{\norm{ \mathbf{\hat{X}} - \mathbf{\bar{X}} }_{2}}{\norm{\mathbf{\bar{X}}}_{2}}.
\end{align}

\subsection{3D Riemann Problems and Characteristic Waves}\label{sec:riemann}

\begin{figure*}
\begin{center}
\centerline{\includegraphics[width=0.8\textwidth]{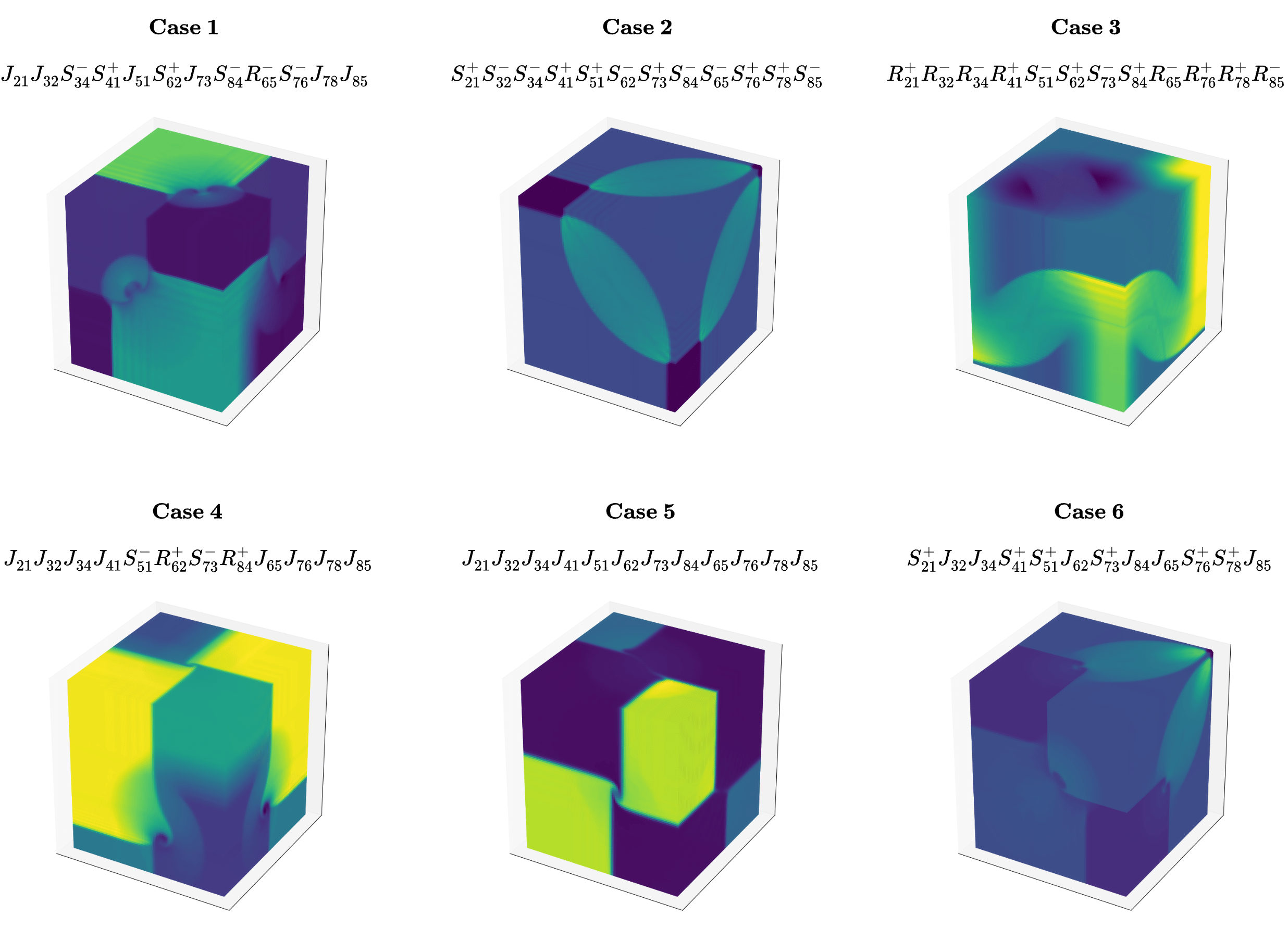}}
\caption{Density fields of the 3D Riemann configurations that we explore in this study.}
\label{fig:cases}
\end{center}
\vskip -0.3in
\end{figure*}

We are interested in solutions of the 3D Euler equations where the initial conditions are described by Riemann problems. As such we briefly discuss this class of problems. For a more in-depth discussion of Riemann problems, refer to \cite{schulz1993classification,hoppe2024systematic}.

A unit cube domain, with dimensions $\left[0,1\right]^{3}$, is divided into 8 octants where each octant is uniformly initialized with its own set of $\left(\rho, u, v,w, p\right)^{T}$. The top octants are numbered 1 to 4, anticlockwise, starting from the octant where $x>0.5, y>0.5, z>0.5$. The bottom octants are numbered 5 to 8 anticlockwise, starting from the octant where $x>0.5, y>0.5, z<0.5$. The initial states of the octants are set such that there exists a single characteristic wave at each of the 12 interfaces that separate the octants. The characteristic wave can be either a rarefaction $R$, shock $S$, or a contact discontinuity $J$. We identify these waves by their positions based on octant numbering, i.e., $lr = \{21, 32, 34, 41, 51, 62, 73, 84, 65, 76, 78,85\}$. In the following let $\omega$ denote the velocity normal to an interface, and let $\bm{\omega^{'}}$ denote the tangential velocities.



The relations for rarefaction waves ${R}_{lr}^{\pm}$ are as follows (assuming polytropic flow):
\begin{align}
& \omega_{l} - \omega_{r} = \pm \frac{2\sqrt{\gamma}}{\gamma - 1}\left( \sqrt{\frac{p_l}{\rho_l}} - \sqrt{\frac{p_r}{\rho_r}} \right),\,\,\,\,\,\,\,\,\frac{\rho_l}{\rho_r} = \left(\frac{p_l}{p_r}\right)^{\frac{1}{\gamma}},\\[6pt]
& \bm{\omega^{'}}_{l} = \bm{\omega^{'}}_{r}, \,\,\,\,\,\,\,\, \omega_l < \omega_r, \,\,\,\,\,\,\,\,sgn\left( p_{l} - p_{r}\right) = \mp 1.
\end{align}
The relations for shock waves ${S}_{lr}^{\pm}$ are as follows:
\begin{align}
&\omega_{l} - \omega_{r} = \sqrt{\frac{\left(p_{l} - p_{r}\right)\left(\rho_{l} - \rho_{r}\right)}{\rho_l \rho_r}},\,\,\,\,\,\,\,\, \frac{\rho_{l}}{\rho_{r}} = \frac{\gamma\left(p_{l} + p_{r}\right) + \left(p_{l} - p_{r}\right)}{\gamma\left(p_{l} + p_{r}\right) - \left(p_{l} - p_{r}\right)},\\[6pt]
& \bm{\omega^{'}}_{l} = \bm{\omega^{'}}_{r},\,\,\,\,\,\,\,\, \omega_l > \omega_r,\,\,\,\,\,\,\,\, sgn\left( p_{l} - p_{r}\right) = \pm 1.
\end{align}
For contact discontinuities ${J}_{lr}$, the pressure and normal velocities are constant, i.e., $\omega_{l} = \omega_{r}$, $p_{l} = p_{r}$. Additionally, in polytropic flows, the density $\rho$ varies arbitrarily. The tangential velocities $\bm{\omega^{'}}$ can also vary arbitrarily, in which case a rotation is induced as a slip line \footnote{It is possible to determine directions $\pm\pm$ for contact discontinuities if the tangential velocities differ. However in 3D this will make the notation cumbersome and so we decide to omit directions for contact discontinuities.}.

Riemann problems (2D or 3D) yield complex flow patterns which arise from the interaction of the initialized elementary waves. They are therefore used as canonical problems for testing compressible flow solvers. We argue that they are also a plausible choice for exploring inverse problems of the Euler equations, such as our reconstruction task. The configurations or cases we explore are shown in Figure \ref{fig:cases}, and their initialization values are summarised in \ref{app:init}.

\subsection{GraphGPS Framework}\label{sec:gps}

\begin{figure*}[h]
\vskip 0.2in
\begin{center}
\centerline{\includegraphics[width=1.0\textwidth]{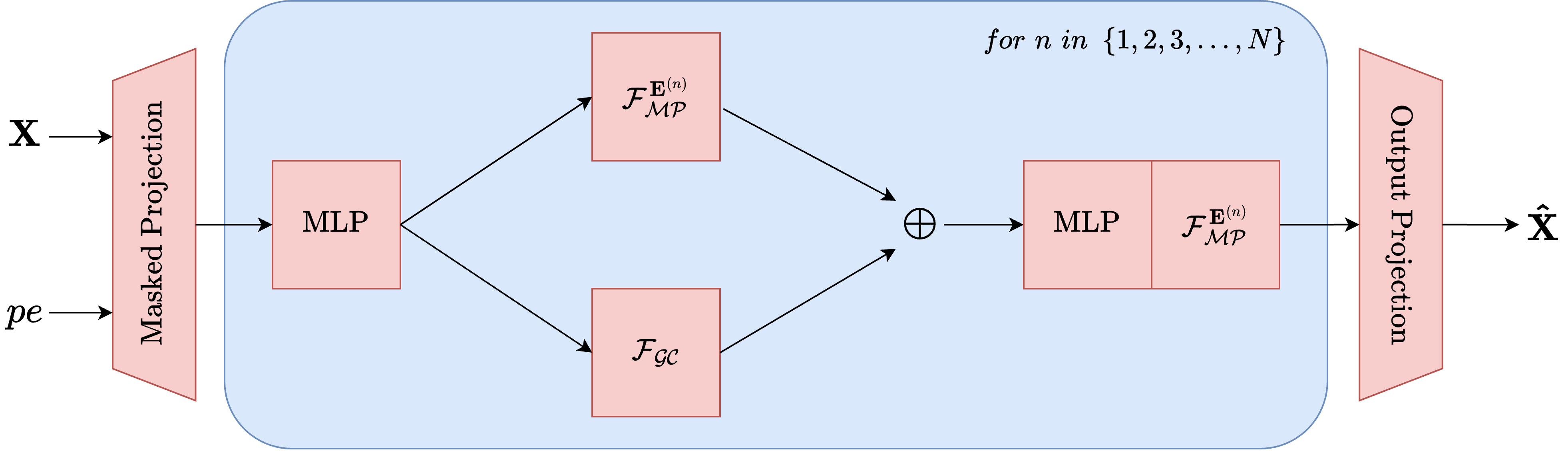}}
\caption{Overview of GraphGPS framework. $\mathcal{F}_{\mathcal{MP}}^{\mathbf{E}^{(n)}}$ represents a message-passing layer acting on edge indices $\mathbf{E}^{(n)}$ for layer $n$. $\mathcal{F}_{\mathcal{GC}}$ denotes a layer that captures global dependencies of its input features. $pe$ denotes positional encodings. Residual connections and normalization layers are omitted for clarity.}
\label{fig:gps}
\end{center}
\vskip -0.3in
\end{figure*}

In this subsection we outline our design of model $\mathcal{F}_{\Theta}$, which follows the GraphGPS framework \cite{rampavsek2022recipe}. Figure \ref{fig:gps} provides a high-level overview of this framework. The framework takes the sparse field $\mathbf{X}$, projects it to higher dimensional space along with positional encodings, then applies $N$ GraphGPS layers before projecting back to the original space. Each GraphGPS layer involves contributions from two modules: $\mathcal{F}_{\mathcal{MP}}$ and $\mathcal{F}_{\mathcal{GC}}$. $\mathcal{F}_{\mathcal{MP}}$ is the message-passing module which captures information at a local level, in the neighbourhood of each node, where the connections between the nodes are defined by edge indices $\mathbf{E}^{(n)}$ for layer $n$.  $\mathcal{F}_{\mathcal{GC}}$ is the module used for computing global context between distant nodes. There are also MLP layers, each involving a linear-ReLU-linear sequence, for processing the combined contributions of $\mathcal{F}_{\mathcal{MP}}$ and $\mathcal{F}_{\mathcal{GC}}$, as well as for mixing node features with node positional encodings. The number of layers $N$ is determined by the size of the set of edge indices, i.e., $\cup_{n=1}^{N}\mathbf{E}^{(n)}$. The set of edge indices is determined from the mask $\mathbf{m}$ via guided message-passing, which we explain next.

\subsubsection{Guided Message-Passing}\label{sec:gmp}

Guided message-passing is underpinned by two rules. The first rule is that during message-passing, information flows strictly from known nodes only (to immediate neighbours). The second rule is that an unknown node becomes 'known' if information has been propagated to it at least once. Formally:

\textbf{Definition} (Guided Message-Passing) : \textit{Let $\mathbf{m} = \mathbf{m}^{(0)} \in \left\{0,1\right\}^{N_{xyz}}$ denote the initial mask.
Let $d_{1}\left(i,j\right)$ denote the $l_1$ distance between node $i$ and node $j$ in their Euclidean space. The set of edge indices $\cup_{n=1}^{N}\mathbf{E}^{(n)}$ can be iteratively generated starting from $\mathbf{m}^{(0)}$ based on the following two rules:}
\begin{align}
    &\mathbf{E}^{(n)} = \left\{ 
    \left(i,j \right) \,\, \mid \,\, d_{1}\left(i,j\right) \in \mathbb{Z^{+}} \cap \left[1,\lambda\right],  \,\,\, m^{(n-1)}_{i} = 1, \,\,\,   \forall i
    \right\}, \,\,\, \lambda \in \left\{1,2,3\right\}, \\[6pt]
    &{m}^{(n)}_{j} = {m}^{(n-1)}_{j} \lor \left( \bigvee_{(i,j) \in \mathbf{E}^{(n)}} {m}_{i}^{(n-1)} \right),
\end{align}

\textit{where $\lor$ denotes the logical OR operator, and $\bigvee$ denotes the logical OR over a set.}

\begin{figure*}[h]
\begin{center}
\centerline{\includegraphics[width=1.0\textwidth]{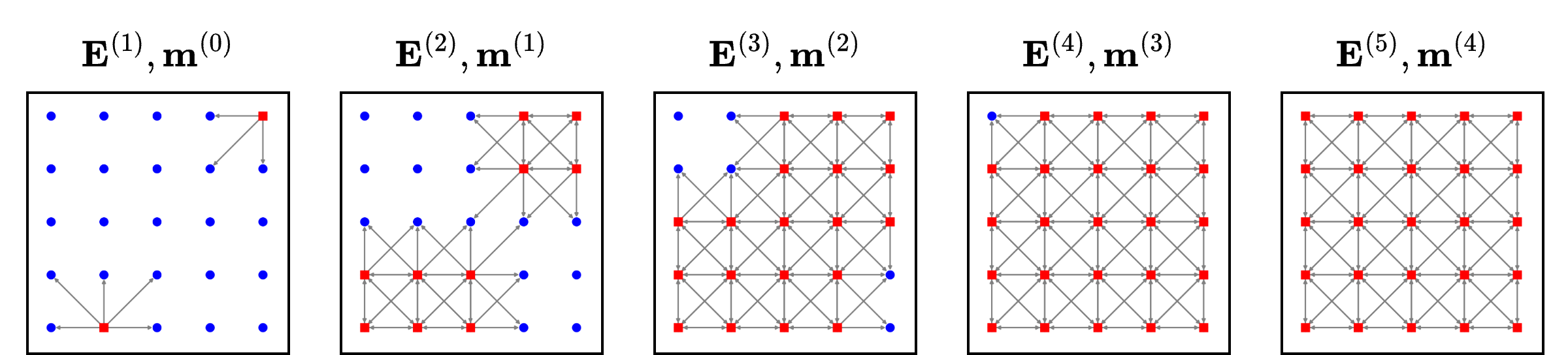}}
\caption{Illustration of guided message-passing. Red squares indicate known nodes. Blue circles indicate unknown nodes.}
\label{fig:cmp}
\end{center}
\vskip -0.3in
\end{figure*}

Parameter $\lambda$ determines whether to consider orthogonal immediate neighbours only ($\lambda=1$), orthogonal + diagonal immediate neighbours ($\lambda=2$), or orthogonal + diagonal + corner immediate neighbours ($\lambda=3$), in determining the next set of edge connections. Figure \ref{fig:cmp} illustrates guided message-passing in 2D with $\lambda=2$.

\subsubsection{Masked Projection and Positional Encoding}\label{sec:maskedproj}

We use masked linear transformations to project input $\mathbf{X}$ to a higher dimensional space, such that in the projected space, the representation of unknown nodes is zeroed. We concatenate these projected features with the linear transformation of Laplacian positional encodings of $\mathbf{X}$ \cite{dwivedi2020generalization,dwivedi2023benchmarking}. Through this masked projection, we ensure that the initially unknown nodes contribute to subsequent layers only through their positional information. This masked projection also ensures the model is agnostic to the padding value used for unknown nodes in $\mathbf{X}$ (should they be non-zero). 

\subsubsection{Candidates for $\mathcal{F}_{\mathcal{MP}}$}\label{sec:mp}

This section outlines the message-passing modules we will explore. One popular and foundational message-passing layer is the graph convolutional layer (GCNConv). For node $i$, GCNConv is expressed as \cite{kipf2016semi}:
\begin{equation}
    \mathbf{h}_i^{\text{out}} = \sigma \left( \sum_{j \in \mathcal{N}(i) \cup \{i\}} \mathbf{W} \mathbf{h}_j \right),
\end{equation}
where $\mathbf{h}$ represents an intermediate feature, $\mathbf{W}$ is the layer weight, $\sigma$ is the ReLU activation function, and $\mathcal{N}\left(i\right)$ denotes the neighbourhood of node $i$ (according to edge indices $\mathbf{E}$). A graph attention convolutional layer (GATConv) is an extension on GCNConv through the introduction of attention weights $\alpha_{ij}$ \cite{velivckovic2017graph}:
\begin{equation}
    \mathbf{h}_i^{\text{out}} = \sigma \left( \sum_{j \in \mathcal{N}(i) \cup \{i\}} \alpha_{ij} \mathbf{W} \mathbf{h}_j \right),
\end{equation}
where the attention weights \( \alpha_{ij} \) are computed as:
\begin{equation}
    \alpha_{ij} = \frac{\exp\left( \text{LeakyReLU} \left( \mathbf{a}^T_{\text{src}} \mathbf{W}\mathbf{h}_i + \mathbf{a}^T_{\text{trg}}\mathbf{W}\mathbf{h}_j \right) \right)}
    {\sum_{k \in \mathcal{N}(i) \cup \{i\}} \exp\left( \text{LeakyReLU} \left( \mathbf{a}^T_{\text{src}} \mathbf{W}\mathbf{h}_i + \mathbf{a}^T_{\text{trg}}\mathbf{W}\mathbf{h}_k \right) \right)},
\end{equation}
and $\mathbf{a}_{\text{src}}^{T}$, $\mathbf{a}_{\text{trg}}^{T}$ are learnable parameters pertaining to the source ($i$) and target ($j$) nodes. In this paper we propose to modify the GATConv attention mechanism by introducing additional parameters $\mathbf{\Tilde{W}}$, $\mathbf{b}_{\text{src}}^{T}$, $\mathbf{b}_{\text{trg}}^{T}$ acting on a representation $\mathbf{Z}$ which encapsulates information about flow discontinuities:
\begin{equation}
    \alpha_{ij} = \frac{\exp\left( \text{LeakyReLU} \left( \mathbf{a}^T_{\text{src}} \mathbf{W}\mathbf{h}_i + \mathbf{a}^T_{\text{trg}}\mathbf{W}\mathbf{h}_j + \mathbf{b}^T_{\text{src}} \mathbf{\Tilde{W}}\mathbf{Z}_i + \mathbf{b}^T_{\text{trg}}\mathbf{\Tilde{W}}\mathbf{Z}_j\right) \right)}
    {\sum_{k \in \mathcal{N}(i) \cup \{i\}} \exp\left( \text{LeakyReLU} \left( \mathbf{a}^T_{\text{src}} \mathbf{W}\mathbf{h}_i + \mathbf{a}^T_{\text{trg}}\mathbf{W}\mathbf{h}_k + \mathbf{b}^T_{\text{src}} \mathbf{\Tilde{W}}\mathbf{Z}_i + \mathbf{b}^T_{\text{trg}}\mathbf{\Tilde{W}}\mathbf{Z}_k \right) \right)},
\end{equation}
where $\mathbf{Z}\in \mathbb{R}^{N_{xyz}\times3N_{t}}$ is obtained from the reconstruction $\mathbf{\hat{X}}$ at the previous training epoch by stacking the following physical quantities:
\begin{align}
    \mathbf{Z} = \left\{\norm{\nabla \rho}_2\in \mathbb{R}^{N_{xyz}\times N_{t}}, \,\,\,\,\,\,  \norm{\nabla p}_2\in \mathbb{R}^{N_{xyz}\times N_{t}}, \,\,\,\,\,\,  \text{max}\left(0,-\text{div}(u,v,w)\right)\in \mathbb{R}^{N_{xyz}\times N_{t}}\right\}.
\end{align}
The density gradient alone is able to capture information on both shocks and contact discontinuities indiscriminately, since they both involve density jumps. The combination of density and pressure gradients however enables the attention mechanism to implicitly distinguish these two wave types since entropy $s = \text{ln}\left(p/\rho^{\gamma}\right)$ increases in the case of shocks. Lastly, the divergence of velocity is present to enable explicit identification of shocks by finding compressible regions. We refer to our modified, shock-aware GATConv as SA-GATConv.

The final message passing layer we explore is SAGEConv which aggregates over a sampled subset of a node's neighbourhood, enabling scalability \cite{hamilton2017inductive}:
\begin{equation}
    \mathbf{h}_i^{\text{out}} = \sigma \left( \mathbf{W}_{1}\mathbf{h}_{i} +  \mathbf{W}_{2}\cdot \text{aggregate} \left( \{ \mathbf{h}_j, \,\,\, \forall j \in \mathcal{N}_{\text{samp}}(i) \subseteq \mathcal{N}(i) \} \right) \right),
\end{equation}
where the aggregation function is either mean, max-pool, or LSTM-based. 

\subsubsection{Candidates for $\mathcal{F}_{\mathcal{GC}}$}\label{sec:gc}

\begin{figure*}
\begin{center}
\centerline{\includegraphics[width=0.5\textwidth]{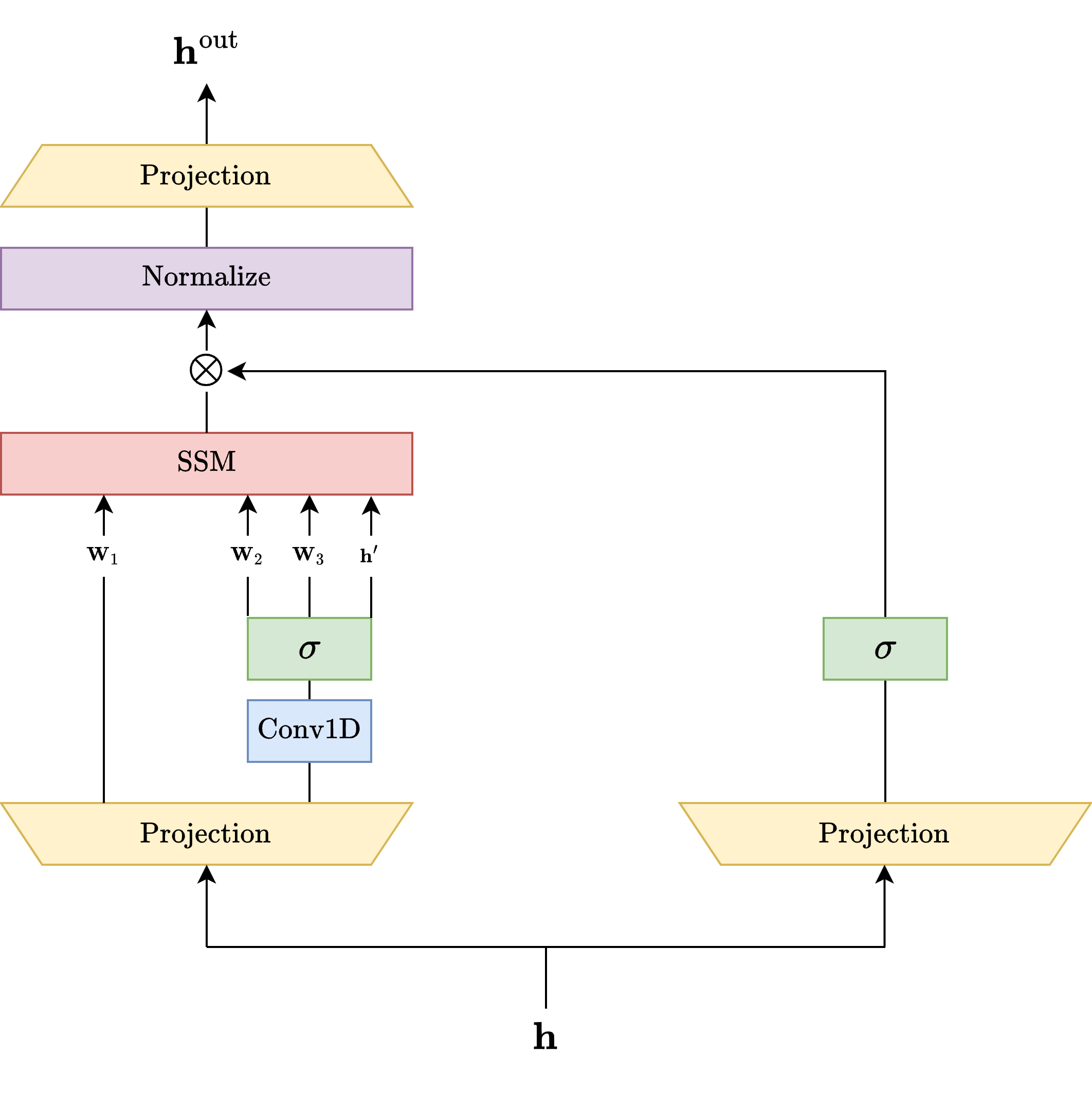}}
\caption{Mamba-2 module.}
\label{fig:mamba2}
\end{center}
\vskip -0.3in
\end{figure*}

We now explore candidates for $\mathcal{F}_{\mathcal{GC}}$, the module for capturing global context. A popular choice for capturing global context is the scaled dot-product attention proposed by \citet{vaswani2017attention}:
\begin{equation}
    \mathbf{h}^{\text{out}} = \text{softmax}\left(\frac{\mathbf{Q}\mathbf{K}^{T}}{\sqrt{d}}\right)\mathbf{V},
\end{equation}
where $\mathbf{Q}$, $\mathbf{K}$, and $\mathbf{V}$ are the queries, keys and values given by $\mathbf{Q} = \mathbf{h}\mathbf{W}_{\mathbf{Q}}$, $\mathbf{K} = \mathbf{h}\mathbf{W}_{\mathbf{K}}$, and $\mathbf{V} = \mathbf{h}\mathbf{W}_{\mathbf{V}}$, and $d$ is the number of hidden channels. This form of attention is however quadratic in complexity and therefore unsuitable for large graphs. A linear-complexity alternative proposed by \citet{katharopoulos2020transformers} aims to approximate the softmax function using kernel functions $\phi\left(\cdot\right)$:
\begin{equation}
    \mathbf{h}^{\text{out}} = \frac{\left(\phi\left(\mathbf{Q}\right)\left(\phi\left(\mathbf{K}\right)^{T}\mathbf{V}\right)\right)}{\left(\phi\left(\mathbf{Q}\right)\left(\phi\left(\mathbf{K}\right)^{T}\mathbf{1}\right)\right)},
\end{equation}
where the kernel function could be exponential, sigmoid, ReLU, ELU + 1, etc. Another way of capturing global context is the Exphormer attention, which is also linear in complexity and takes a graph-centric approach. If we view $\mathbf{h}$ graphically, then for node $i$ (corresponding to $\mathbf{h}_i$) the attention (and its propagation) are computed as \cite{shirzad2023exphormer}:
\begin{equation}
    \mathbf{h}_{i}^{\text{out}} = \frac{\sum_{j \in \mathcal{N}_{\text{eg}}(i)} \mathbf{O}_{ij} \mathbf{V}_j}
    {\sum_{j \in \mathcal{N}_{\text{eg}}(i)} \mathbf{O}_{ij}}, \,\,\,\,\, \mathbf{O}_{ij} = \exp \left( \frac{\mathbf{Q}_i  \mathbf{K}_j}{\sqrt{d}} \right),
\end{equation}
where $\mathcal{N}_{\text{eg}}(i)$ denotes the neighbours of node $i$ as defined by an expander graph of $\mathbf{h}$. Expander graphs are sparse approximations of complete graphs which introduce edge connections between nodes that are distant in the original graph. For details on how expander graphs are generated, we refer the reader to \cite{shirzad2023exphormer}. The expander graph is usually complemented with a small number of virtual nodes which form edge connections with all the nodes of the expander graph. These global nodes allow for communication between nodes that are not directly connected in the expander graph, introducing another flavour of global attention.

The final form of $\mathcal{F}_{\mathcal{GC}}$ we explore is the Mamba-2 module, which is best explained schematically as in Figure \ref{fig:mamba2}. Mamba-2 does not explicitly compute attention scores to capture global context. Rather, it relies on a gating mechanism and a selective state space model to determine which information to discard and retain. The state space model updates the intermediate input $\mathbf{h}'$ through a fixed size latent variable $\mathbf{\Psi}$, and its selectivity arises from making the weights $\mathbf{W}_1$, $\mathbf{W}_2$, and $\mathbf{W}_3$ themselves input-dependent \cite{gu2023mamba,dao2024transformers}. If we view the intermediate input as a sequence (rather than as a graph) indexed by $n$, then the recurrent state-space model is defined as:
\begin{align}
    &\mathbf{\Psi}^{(n)} = \mathbf{W}_{1}\mathbf{\Psi}^{(n-1)} + \mathbf{W}_{2}\mathbf{h}'^{(n)}, \\
    & \mathbf{y}^{(n)} = \mathbf{W}_3 \mathbf{\Psi}^{(n)}.
\end{align}
Mamba-2 is linear in complexity and has been proposed as an alternative to the transformer model. We remark that a crucial feature of Mamba-2 is that weight $\mathbf{W}_1$ has a \textit{scalar times identity matrix} structure. By imposing this structure on $\mathbf{W}_1$, a duality can be derived between state space models and attention, enabling the design of an efficient algorithm for computing the state space model which has linear complexity whilst making use of the GPU-friendly matrix multiplications of attention \cite{dao2024transformers}.

\subsection{Multi-Loss Function}

To learn the weights of function $\mathcal{F}_{\Theta}$, we optimize a data loss $\mathcal{L}_{\text{data}}$ and a physical loss $\mathcal{L}_{\text{phy}}$ with respect to the weights. The data loss is computed only for the initially known nodes as defined by $\mathbf{m} = \mathbf{m}^{(0)}$:
\begin{align}
\mathcal{L}_{\text{data}} = \frac{1}{N_{tc}\sum \mathbf{m}}\norm{\mathbf{m}\odot\left(\mathbf{X} - \hat{\mathbf{X}}\right)}_{2}^{2}.
\end{align}
For the physical loss, we first map reconstruction $\mathbf{\hat{X}}$ back to conserved variables $\mathbf{\hat{U}}\in \mathbb{R}^{N_{t}\times N_{c} \times N_{x} \times N_{y} \times N_{z}}$, which we then use to compute a Godunov loss function\footnote{We use here notation $\left(i,j,k\right)$ to denote spatial indices in 3D. This notation is not to be confused with earlier sections where notation $\left(i,j\right)$ is used to denote graph nodes.}:
\begin{align}\label{eq:godloss}
    &\mathcal{L}_{\text{phy}} =  \frac{1}{N} \sum_{c}\sum_{t}\sum_{i,\,j,\,k} \left( 
    \mathbf{\hat{U}}_{i,\,j,\,k}^{(t+1,\,c)} - \mathbf{\hat{U}}_{i,\,j,\,k}^{(t,\,c)}  + \lambda_{x}\left[\Delta{\mathbf{F}}_{i,\,j,\,k}^{(t,\,c)} \right] 
    + \lambda_{y}\left[\Delta{\mathbf{G}}_{i,\,j,\,k}^{(t,\,c)}\right]
    + \lambda_{z}\left[\Delta{\mathbf{H}}_{i,\,j,\,k}^{(t, \, c)}\right]\right)^{2},\\
    &\Delta{\mathbf{F}}_{i,\,j,\,k}^{(t,\,c)} = {\mathbf{F}}_{i + \frac{1}{2},\,j,\,k}^{(t,\,c)} - {\mathbf{F}}_{i - \frac{1}{2},\,j,\,k}^{(t,\,c)}, \\[8pt]
    &\Delta{\mathbf{G}}_{i,\,j,\,k}^{(t,\,c)} = {\mathbf{G}}_{i,\,j + \frac{1}{2},\,k}^{(t,\,c)} - {\mathbf{G}}_{i,\,j - \frac{1}{2}\,,k}^{(t,\, c)},
    \\[8pt]
    &\Delta{\mathbf{H}}_{i,\,j,\,k}^{(t,\,c)} = {\mathbf{H}}_{i,\,j,\,k+\frac{1}{2}}^{(t,\,c)} - {\mathbf{H}}_{i,\,j,\,k-\frac{1}{2}}^{(t,\,c)},
\end{align}

where $\lambda_{x} = \Delta t/\Delta x$, $\lambda_{y} = \Delta t/\Delta y$, $N = N_{xyz}N_{tc}$, and $\Delta{\mathbf{F}}$, $\Delta{\mathbf{G}}$, $\Delta{\mathbf{H}}$ are the net $x,y,z$ fluxes computed from $\mathbf{\hat{U}}$ using a Riemann solver. For the purposes of this paper we use the HLLC Riemann solver to estimate the fluxes, which assumes a three-wave structure at each cell interface, as outlined in \ref{app:hllc}. We refer the reader to \cite{cassia2025godunov} for more information on Godunov losses and the justification of their use for modeling hyperbolic conservation laws.

Rather than optimizing the summation $\mathcal{L}_{\text{data}} + \mathcal{L}_{\text{phy}}$ with one Adam optimizer, we choose to optimize the two losses separately using two separate Adam optimizers. Therefore, within each epoch, two back-propagations are applied, and two updates are made to the model weights. Despite the increased computational overhead, we find that this setup significantly stabilizes training.


%% file: Experiments.tex
\section{Experiments}\label{sec:exp}

This section is divided into three parts. Firstly, we perform an ablation study on the message-passing and global attention components of the GraphGPS framework. Secondly, we examine the performance of guided-message passing compared to standard, dense message-passing under different node connectivities and initializations. Thirdly, we examine performance of our GraphGPS model against benchmark ML models.

All models examined are trained to 5000 epochs using an Adam optimizer, with a learning rate of $10^{-4}$, on a NVIDIA A100 GPU. A model is trained for each flow case outlined in Section \ref{sec:riemann}. The number of hidden channels (or projected dimension) is set to 64 for all models. We explore a grid of spatial resolution $64^3$ and time resolution $5\times10^{-4}$. For visualization, we find that plotting diagonal slices through the Riemann cube, defined by normal vectors $\left(\pm1,\pm1,\pm1\right)^{T}$, is most effective for revealing the 3D nature of the flow.

\begin{figure*}[h!]
\begin{center}
\centerline{\includegraphics[width=0.9\textwidth]{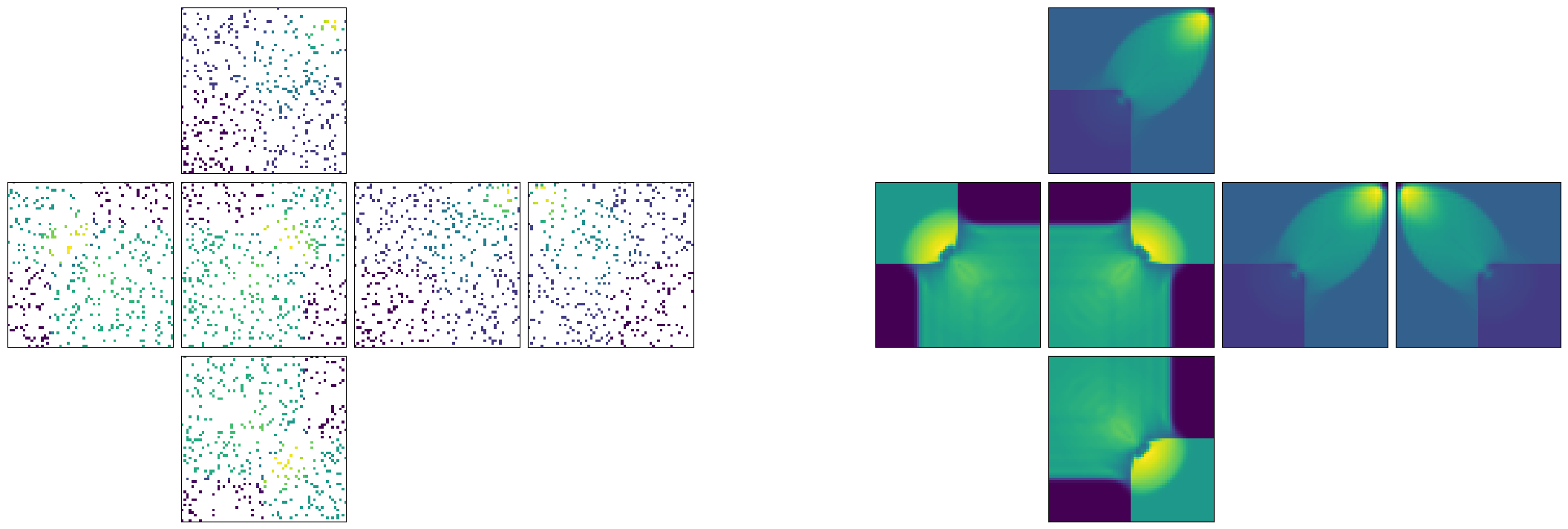}}
\caption{Distribution of known points on the faces of Case 6 (left), and the corresponding true field for Case 6 (right). The other cases have the same distribution of known points. The density field is shown.}
\label{fig:sensors}
\end{center}
\vskip -0.3in
\end{figure*}

Figure \ref{fig:sensors} shows the distribution of observed points on the faces of the Riemann cube, to give a sense of the level of sparsity we explore. In tangible terms, 90\% of points are unknown. All cases explored have the same distribution of known points.

\subsection{Ablation Study}

For all experiments explored in this subsection, we use guided message-passing with $\lambda = 3$ and masked projection, which are outlined in Section \ref{sec:gmp} and Section \ref{sec:maskedproj}, respectively.

\begin{figure*}[h!]
\begin{center}
\centerline{\includegraphics[width=0.9\textwidth]{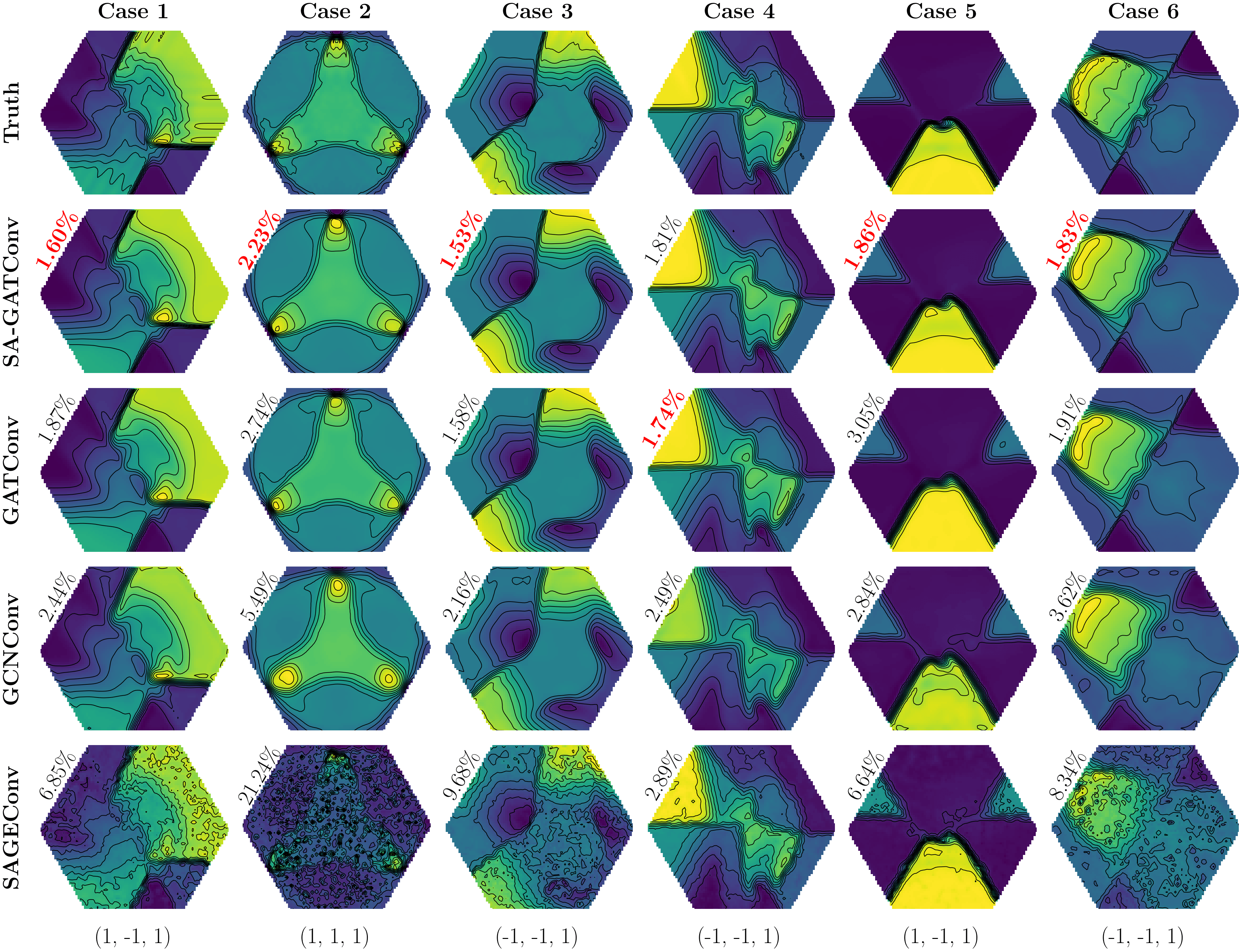}}
\caption{Performance of different $\mathcal{F}_{\mathcal{MP}}$ candidates when combined with Mamba-2, along with the use of masked projection and guided message-passing $(\lambda=3)$.  The relative L2 error across all physical channels in $\mathbf{X}$ is indicated on the top left of each subplot, and the density field is plotted with values normalized between [0,1]. The normal vectors that define the planes of view are indicated at the bottom for each case. The full 3D density fields for the cases are shown in Figure \ref{fig:cases}.}
\label{fig:A_outputs}
\end{center}
\vskip -0.3in
\end{figure*}

We first examine the effect of using different message-passing layers, $\mathcal{F}_{\mathcal{MP}}$, on the performance of the framework in Figure \ref{fig:gps}, whist keeping $\mathcal{F}_{\mathcal{GC}}$ the same (Mamba-2). We examine the performance when using GATConv, our SA-GATConv, GCNConv, and SAGEConv. See Section \ref{sec:mp} for an overview of these layers. For GATConv and SA-GATConv, we use one attention head. For SAGEConv we use max pooling as the aggregation method. Figure \ref{fig:A_outputs} shows the density plots and error for each message-passing layer and for each case. We find that SA-GATConv and GATConv give the most faithful reconstructions both visually and numerically. On the other hand, the susceptibility of GCNConv to oversmoothing can be observed where there are shocks and slip lines. SAGEConv performs the worst and gives noisy reconstructions. Table \ref{tab:A_compute} compares the computational expense of GraphGPS when using the different message-passing layers. Due to their attention mechanisms, GATConv and SA-GATConv require more memory and time. Furthermore the modified attention mechanism of SA-GATConv slightly increases the overhead compared to GATConv.

\begin{table}[h!]
\caption{Compute information of different $\mathcal{F}_{\mathcal{MP}}$ candidates when combined with Mamba-2, along with the use of masked projection and guided message-passing $(\lambda=3)$.}
    \centering
    \setlength{\tabcolsep}{6pt}
    \begin{small}
    \begin{tabular}{lcccc}
        \toprule
        & \textbf{SA-GATConv} & \textbf{GATConv} & \textbf{GCNConv} & \textbf{SAGEConv} \\
        \midrule
        Max Memory Allocated (GB) & 37.70 & 36.90 & 15.40 & 15.10 \\
        Avg Time per Epoch (s)              & 1.64  & 1.60  & 1.43  & 1.40  \\
        Total Training Time (hours)        & 2.28  & 2.22  & 1.99  & 1.95  \\
        \bottomrule
    \end{tabular}
    \label{tab:A_compute}
    \end{small}
\end{table}

Examining Figure \ref{fig:A_outputs} for the top performers in more detail, we find that SA-GATConv yields slightly lower errors than GATConv. It is also interesting to examine the physical features of the cases in more detail for these two message-passing layers, which appear sharper for SA-GATConv. For Case 1, we observe this in the shock and slip line towards the bottom right. For Case 2, we observe the same at the tips of the arrow-like subsonic region produced from the interaction of the initial shock waves. The same can be observed in Case 3 for the shock that bends towards the center and dissolves into rarefaction fans. For Case 4, this observation can be made for the slip line that bends the rarefaction, as well as for the bow-shock on the bottom right. In Case 5, the bent slip lines have a more pronounced shape for SA-GATConv that better match the truth compared to GATConv. Finally, Case 6 shows the most significant difference in shock definition between the two message-passing layers when observing the two shock fronts towards the left of the density plots.

\begin{figure*}[h!]
\begin{center}
\centerline{\includegraphics[width=0.9\textwidth]{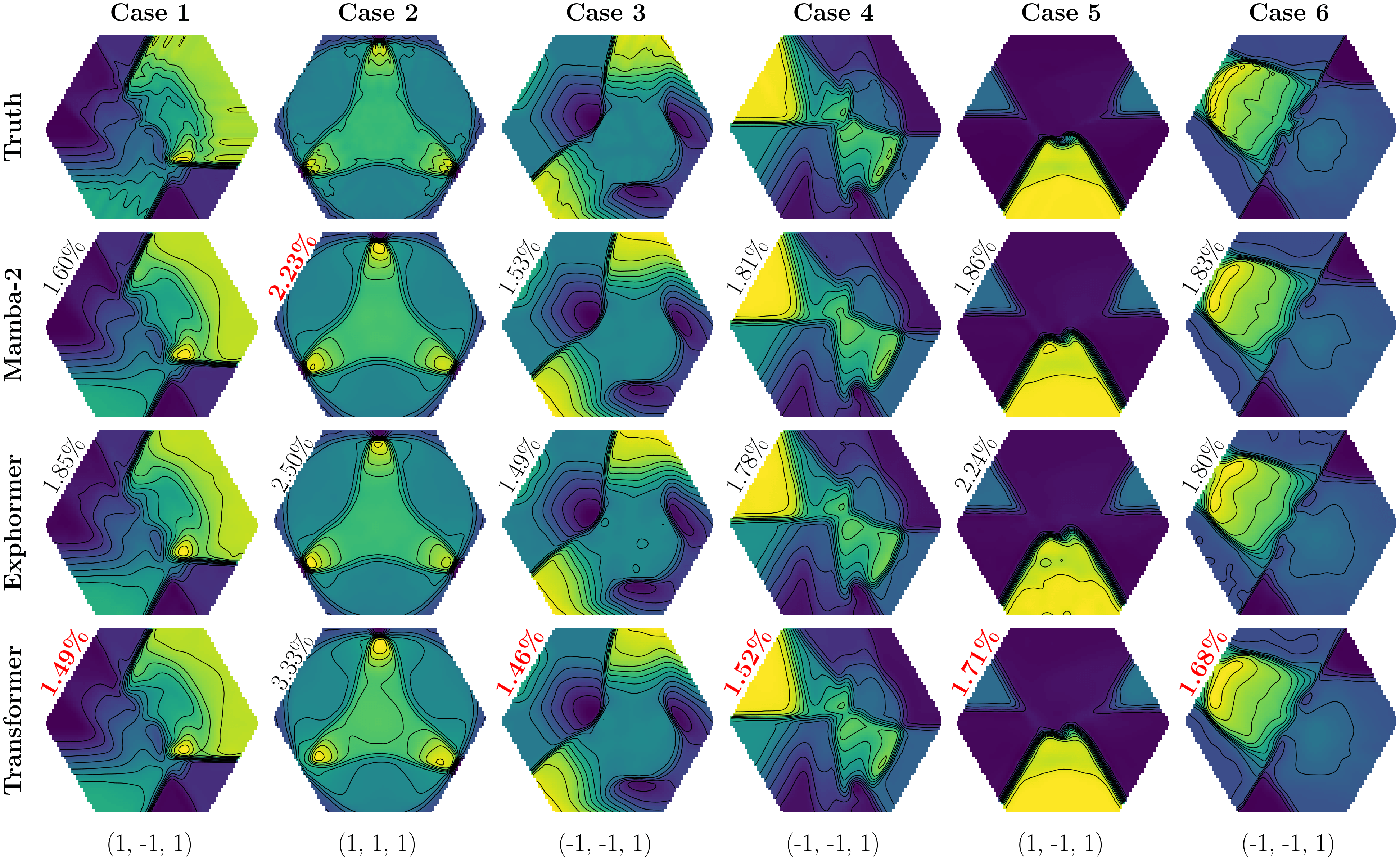}}
\caption{Performance of different $\mathcal{F}_{\mathcal{GC}}$ candidates when combined with SA-GATConv, along with the use of masked projection and guided message-passing $(\lambda=3)$. The relative L2 error across all physical channels in $\mathbf{X}$ is indicated on the top left of each subplot, and the density field is plotted with values normalized between [0,1]. The normal vectors that define the planes of view are indicated at the bottom for each case. The full 3D density fields for the cases are shown in Figure \ref{fig:cases}.}
\label{fig:B_outputs}
\end{center}
\vskip -0.3in
\end{figure*}

Next, we examine the effect of different methods for capturing global context, $\mathcal{F}_{\mathcal{GC}}$, on the performance of the framework in Figure \ref{fig:gps}, whist keeping $\mathcal{F}_{\mathcal{MP}}$ the same (SA-GATConv). We examine the performance when using Mamba-2, linear transformer, and Exphormer. See Section \ref{sec:gc} for an overview of these modules. For all three modules we use 16 heads. For the linear transformer we use kernel function $\phi = \text{ELU}(x) + 1$. For the Exphormer, the degree of the expander graph is 10, and the number of virtual nodes is 5. Examining Figure \ref{fig:B_outputs} we find that the linear transformer marginally outperforms the other two candidates, with the exception of Case 2 where the reconstruction appears least faithful. Of more significance is the computational expense of the three candidates outlined in Table \ref{tab:B_compute}, where Mamba-2 has the least overhead in terms of time and memory. Despite the sparsification of the graph used in Exphormer, the memory consumption is almost double that of the other candidates, due to the propagation step and due to the introduction of additional edge connections through virtual nodes.

\begin{table}[h!]
\caption{Compute information of different $\mathcal{F}_{\mathcal{GC}}$ candidates when combined with SA-GATConv, along with the use of masked projection and guided message-passing $(\lambda=3)$.}
    \centering
    \setlength{\tabcolsep}{6pt}
    \begin{small}
    \begin{tabular}{lccc}
        \toprule
        & \textbf{Mamba-2} & \textbf{Exphormer} & \textbf{Transformer} \\
        \midrule
        Max Memory Allocated (GB) & 37.70 & 77.00 & 37.80 \\
        Avg Time per Epoch (s)              & 1.63  & 2.59  & 2.50  \\
        Total Training Time (hours)        & 2.26  & 3.60  & 3.48  \\
        \bottomrule
    \end{tabular}
    \label{tab:B_compute}
    \end{small}
\end{table}

It is evident from this ablation study that $\mathcal{F}_{\mathcal{MP}}$ is more influential to the performance of GraphGPS compared to $\mathcal{F}_{\mathcal{GC}}$, in the context of our reconstruction task.

\subsection{Guided Message-Passing vs Dense Message-Passing}


For all experiments explored in this subsection, we use SA-GATConv for $\mathcal{F}_{\mathcal{MP}}$ and Mamba-2 for $\mathcal{F}_{\mathcal{GC}}$, which are explained in detail in Section \ref{sec:mp} and Section \ref{sec:gc}, respectively.

\begin{figure*}[h!]
\begin{center}
\centerline{\includegraphics[width=0.9\textwidth]{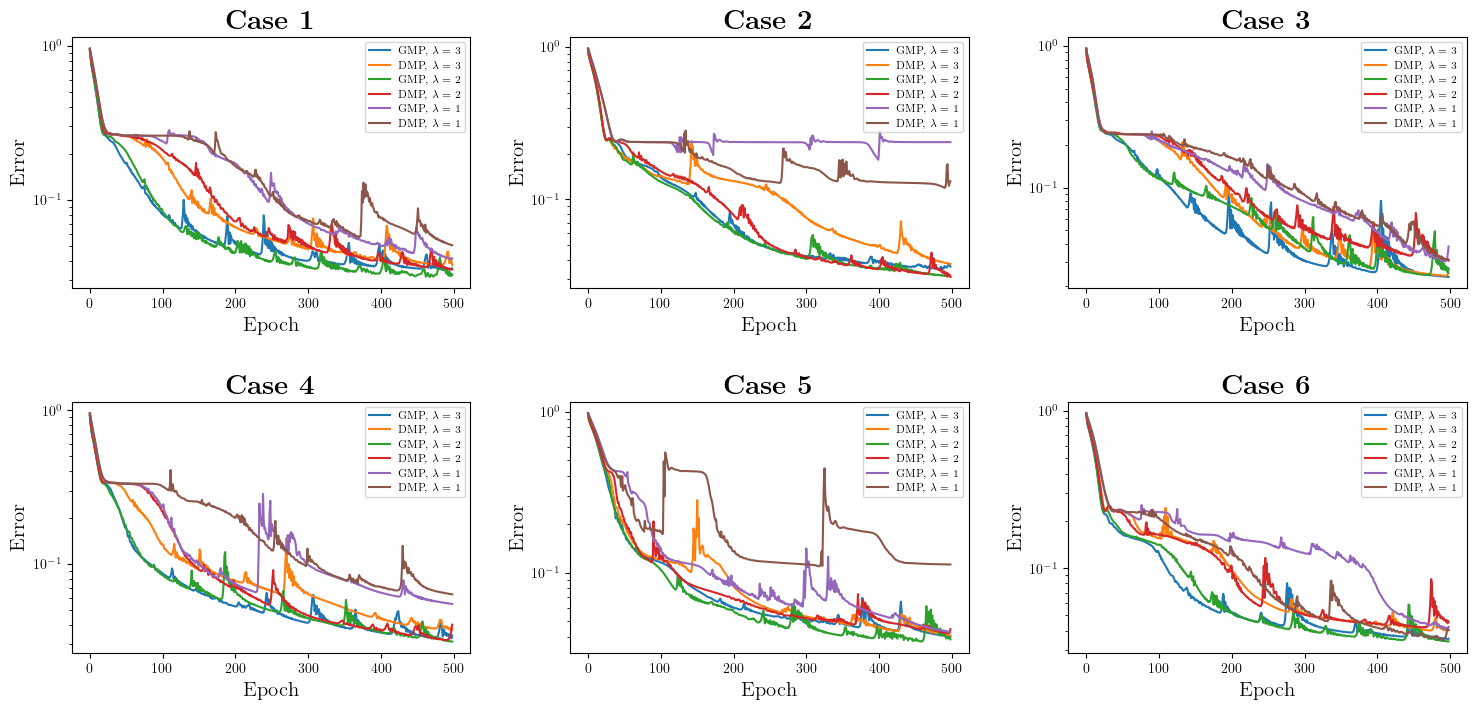}}
\caption{Convergence behaviour of guided/dense message-passing with $\lambda = \left\{1,2,3\right\}$ over the first 500 epochs. We use SA-GATConv and Mamba-2 for $\mathcal{F}_{\mathcal{MP}}$ and $\mathcal{F}_{\mathcal{GC}}$, respectively, and use masked projection.}
\label{fig:D_conv}
\end{center}
\vskip -0.3in
\end{figure*}

We are interested to see how guided message-passing performs in comparison to dense-message passing. An overview of guided message-passing is outlined in Section \ref{sec:gmp}. By dense-message passing, we mean that for all layers of GraphGPS, each node is connected to all its immediate neighbours. Like guided message-passing, the immediate neighbours could be orthogonal neighbours $(\lambda = 1)$, orthogonal + diagonal neighbours $(\lambda = 2)$, or orthogonal + diagonal + corner neighbours $(\lambda = 3)$. It is interesting to compare the performance of guided and dense message-passing under these three different types of edge connectivities, assuming masked projection for all runs. In terms of error we observe only a marginal difference in performance between the message-passing types at $\lambda =3$ and $\lambda=2$, with no significant difference from a visual point of view. We see however that performance significantly degrades for both message-passing types when $\lambda =1$, indicating that the additional diagonal neighbours are critical to the reconstruction process. Furthermore, for $\lambda = 1$, guided message-passing performs significantly worse than dense message-passing for Cases 2, 3, 5, and 6. We thus conclude that in the case of $\lambda=2$ and $\lambda=3$, the use of guided message-passing, which is sparser in terms of number of edge connections, does not come at the cost of performance degradation, as in the case of $\lambda =1$. This sparseness of guided message-passing also results in reduced computational overhead as indicated in Table \ref{tab:D_compute}, especially in terms of memory. Also of note is that the savings also appear less significant going from $\lambda = 3$ to $\lambda = 1$. Another performance gain of guided message-passing appears to be in terms of initial convergence behaviour. Figure \ref{fig:D_conv} plots the evolution of error in the first 500 epochs, where it is evident that guided message-passing at $\lambda = 2$ and $\lambda = 3$ tend towards a stable solution quicker than the other settings, because they do not encounter plateaus as the other settings do. These plateaus are longer and more frequent at $\lambda=1$ for both message-passing types, and are accompanied by sudden jumps in error, which indicate unstable training. This may explain the degraded reconstructions at $\lambda =1$.

\begin{figure*}[h!]
\begin{center}
\centerline{\includegraphics[width=0.9\textwidth]{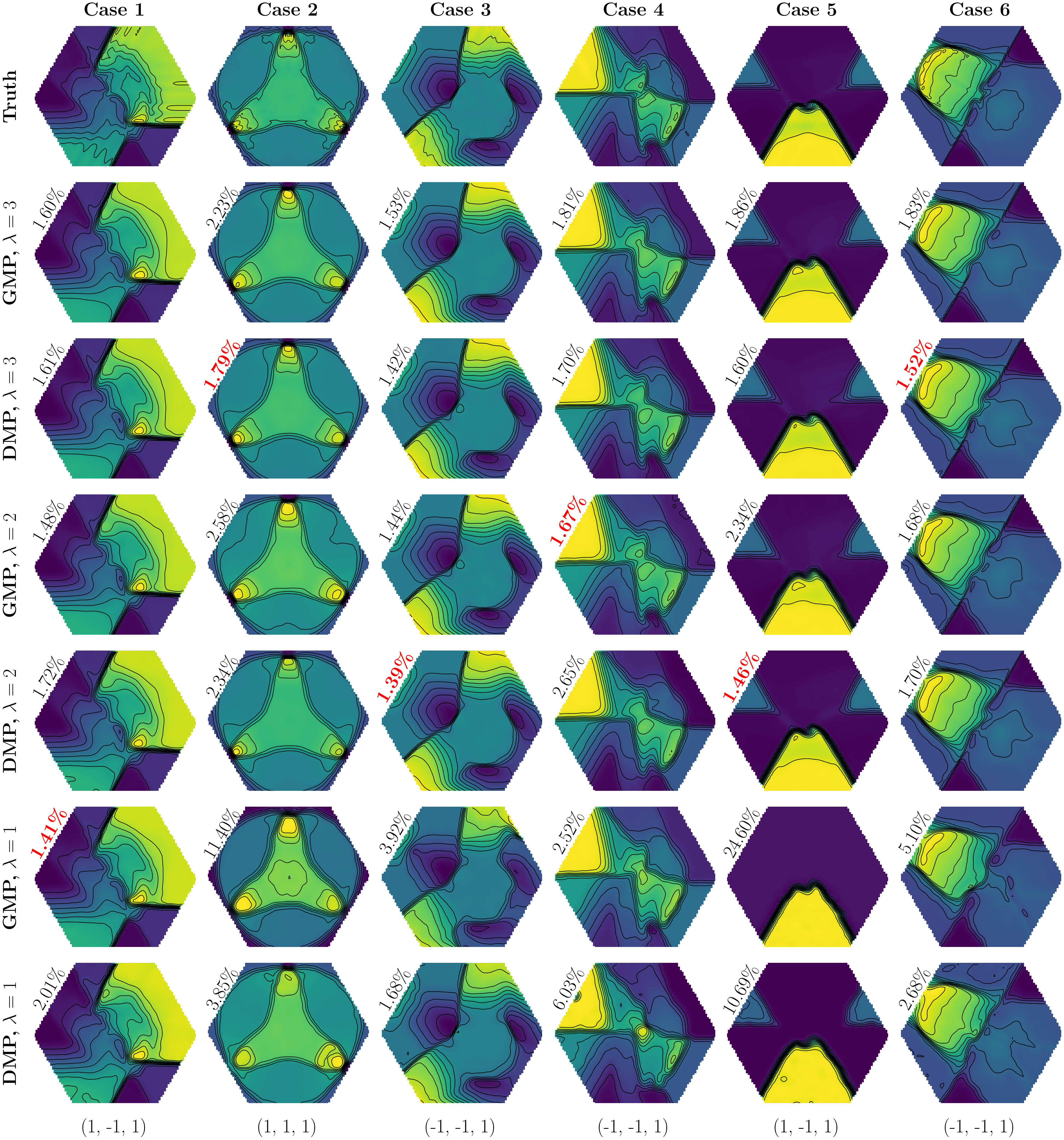}}
\caption{Performance of guided/dense message-passing with $\lambda = \left\{1,2,3\right\}$. We use SA-GATConv and Mamba-2 for $\mathcal{F}_{\mathcal{MP}}$ and $\mathcal{F}_{\mathcal{GC}}$, respectively, and use masked projection. The relative L2 error across all physical channels in $\mathbf{X}$ is indicated on the top left of each subplot, and the density field is plotted with values normalized between [0,1]. The normal vectors that define the planes of view are indicated at the bottom for each case. The full 3D density fields for the cases are shown in Figure \ref{fig:cases}.}
\label{fig:D_outputs}
\end{center}
\vskip -0.3in
\end{figure*}

\begin{table}[h!]
\caption{Compute information of guided/dense message-passing with $\lambda = \left\{1,2,3\right\}$. We use SA-GATConv and Mamba-2 for $\mathcal{F}_{\mathcal{MP}}$ and $\mathcal{F}_{\mathcal{GC}}$, respectively, and use masked projection.}
    \centering
    \setlength{\tabcolsep}{6pt}
    \begin{small}
    \begin{tabular}{lcccccc}
        \toprule
        & \textbf{GMP, $\lambda = 3$} & \textbf{DMP, $\lambda = 3$} & \textbf{GMP, $\lambda = 2$} & \textbf{DMP, $\lambda = 2$} & \textbf{GMP, $\lambda = 1$} & \textbf{DMP, $\lambda = 1$} \\
        \midrule
        Max Memory Allocated (GB) & 37.70 & 41.80 & 29.80 & 32.90 & 28.30 & 29.70 \\
        Avg Time per Epoch (s)              & 1.63  & 1.75  & 1.46  & 1.54  & 1.63  & 1.67  \\
        Total Training Time (hours)        & 2.26  & 2.43  & 2.03  & 2.14  & 2.26  & 2.32  \\
        \bottomrule
    \end{tabular}
    \label{tab:D_compute}
    \end{small}
\end{table}

\begin{figure*}[h!]
\begin{center}
\centerline{\includegraphics[width=0.9\textwidth]{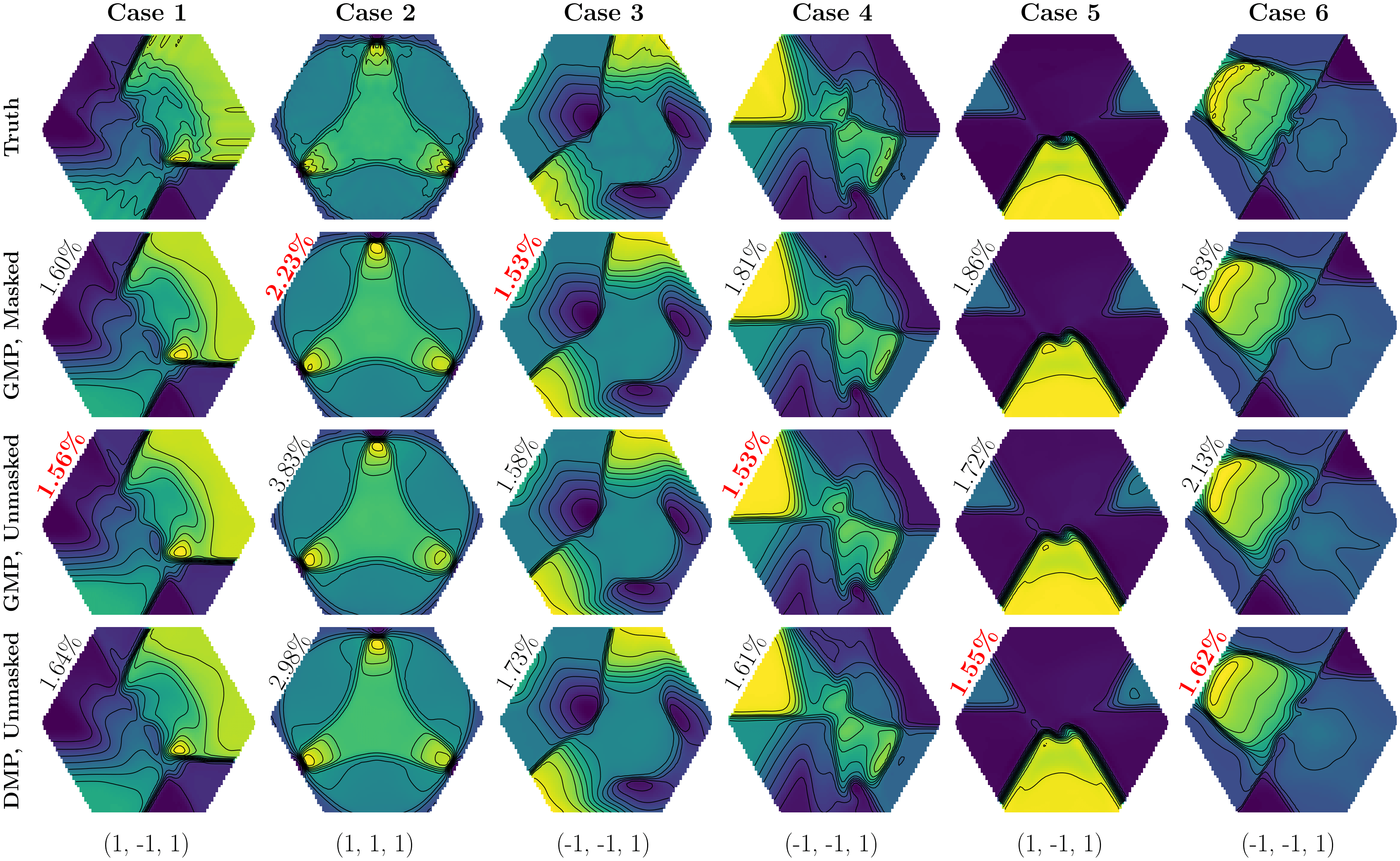}}
\caption{Performance of masked/unmasked projection and guided/dense message-passing $(\lambda = 3)$. We use SA-GATConv and Mamba-2 for $\mathcal{F}_{\mathcal{MP}}$ and $\mathcal{F}_{\mathcal{GC}}$, respectively. The relative L2 error across all physical channels in $\mathbf{X}$ is indicated on the top left of each subplot, and the density field is plotted with values normalized between [0,1]. The normal vectors that define the planes of view are indicated at the bottom for each case. The full 3D density fields for the cases are shown in Figure \ref{fig:cases}.}
\label{fig:C_outputs}
\end{center}
\vskip -0.3in
\end{figure*}

\begin{figure*}[h!]
\vskip 0.2in
\begin{center}
\centerline{\includegraphics[width=0.9\textwidth]{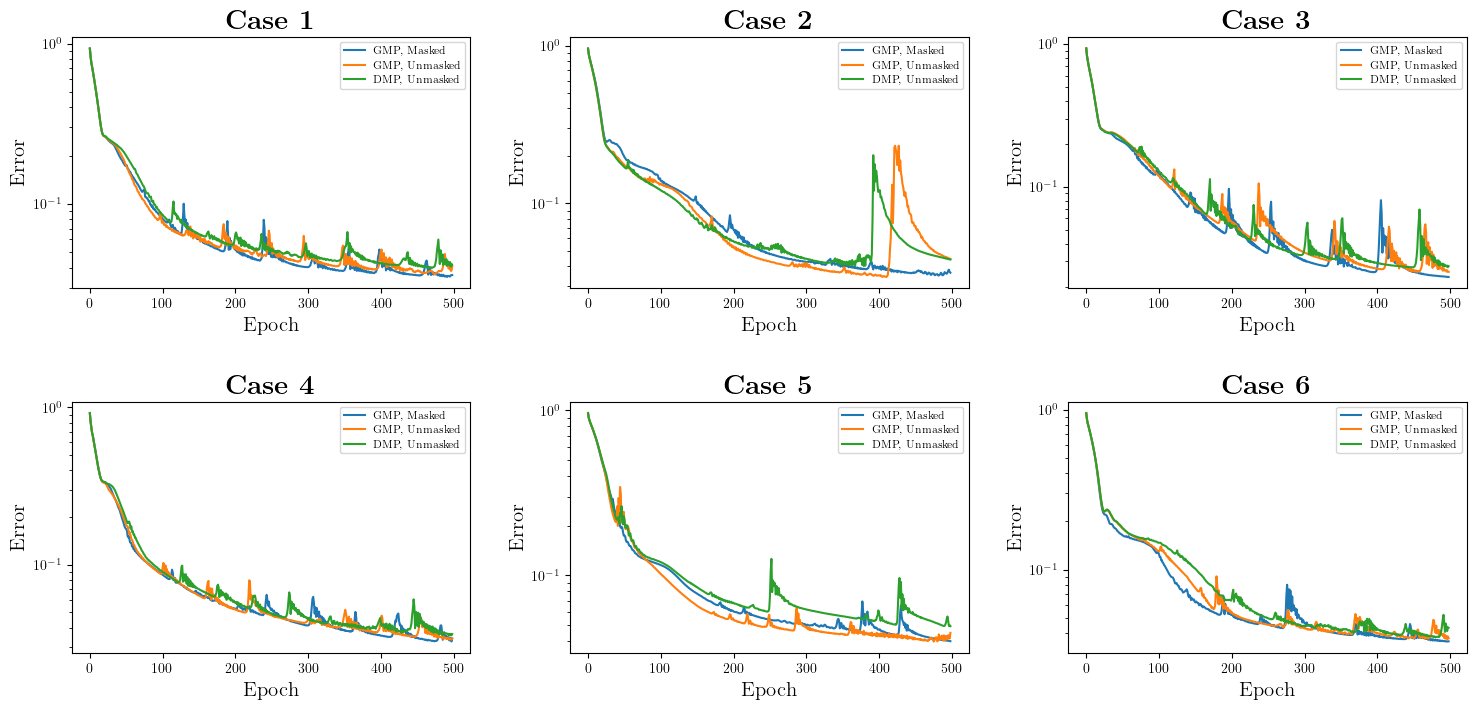}}
\caption{Convergence behaviour of masked/unmasked projection and guided/dense message-passing $(\lambda = 3)$ over the first 500 epochs. We use SA-GATConv and Mamba-2 for $\mathcal{F}_{\mathcal{MP}}$ and $\mathcal{F}_{\mathcal{GC}}$, respectively.}
\label{fig:C_conv}
\end{center}
\vskip -0.3in
\end{figure*}

It is also of interest to compare the two message-passing types when the input projection is masked, or normal (unmasked). Section \ref{sec:maskedproj} outlines masked projection. Where unmasked projection is used, we initialize unknown nodes using feature propagation as is usually done in the literature for graph-based reconstructions \cite{duthe2023graph, danciu2024flow}\footnote{Feature propagation is a pre-processing step where missing features are interpolated from known features using propagation.}. For all runs in Figure \ref{fig:C_outputs} we use $\lambda = 3$. Figures \ref{fig:C_outputs} and \ref{fig:C_conv} suggest there is no notable improvement in accuracy or stability when introducing unmasked projection (with feature propagation) and dense message-passing. This suggests that the zeroed representation of unknown nodes in the projected space when using masked projection is sufficient, and no pre-processing such as feature propagation is required. This also justifies the use of masked projection, which unlike normal projection, is agnostic to the initialization of unknown nodes. For Case 2, one can even observe a degraded reconstruction when using unmasked projection, for both guided and dense message-passing. This appears to coincide with the unstable jumps in error observed in Figure \ref{fig:C_conv} for Case 2, which is not encountered for masked, guided message-passing.

\subsection{Comparison with Benchmarks}

\begin{figure*}[h!]
\begin{center}
\centerline{\includegraphics[width=0.9\textwidth]{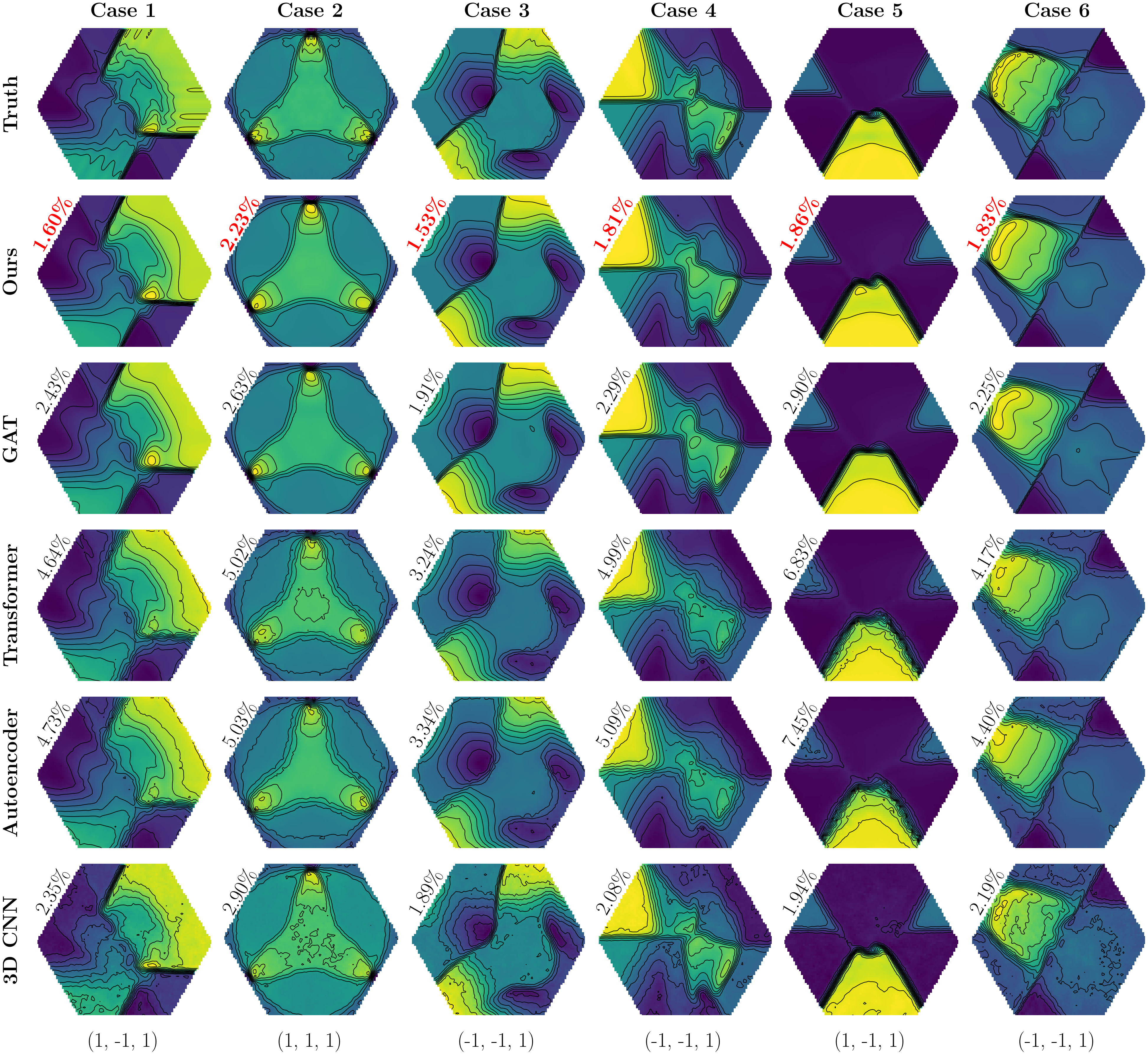}}
\caption{Performance of our architecture against different benchmarks. For our architecture, we use SA-GATConv and Mamba-2 for $\mathcal{F}_{\mathcal{MP}}$ and $\mathcal{F}_{\mathcal{GC}}$, respectively, along with guided message-passing $(\lambda = 3)$ and masked projection. The relative L2 error across all physical channels in $\mathbf{X}$ is indicated on the top left of each subplot, and the density field is plotted with values normalized between [0,1]. The normal vectors that define the planes of view are indicated at the bottom for each case. The full 3D density fields for the cases are shown in Figure \ref{fig:cases}.}
\label{fig:E_outputs}
\end{center}
\vskip -0.3in
\end{figure*}

We compare our GraphGPS architecture with a number of benchmarks, such as graph attention network (GAT), linear transformer, autoencoder, and 3D convolutional neural network (3D CNN). The architectural details of these models are outlined in \ref{app:bm}. For our GraphGPS architecture, we use SA-GATConv and Mamba-2 for $\mathcal{F}_{\mathcal{MP}}$ and $\mathcal{F}_{\mathcal{GC}}$, respectively, along with guided message-passing $(\lambda = 3)$, and masked projection. From Figure \ref{fig:E_outputs}, we find that our model has superior accuracy compared to the benchmarks for all cases, albeit at an increased computational cost as shown in Table \ref{tab:E_compute}. All the benchmarks result in over-smoothing of shocks and sliplines. Moreover, the transformer and autoencoder models produce jagged contours, as does the 3D CNN but to a lesser extent. The 3D CNN also produces noisy artifacts.

\begin{table}[h!]
\caption{Compute information of our architecture against different benchmarks. For our architecture, we use SA-GATConv and Mamba-2 for $\mathcal{F}_{\mathcal{MP}}$ and $\mathcal{F}_{\mathcal{GC}}$, respectively, along with guided message-passing $(\lambda = 3)$ and masked projection.}
    \centering
    \setlength{\tabcolsep}{6pt} 
    \begin{small}
    \begin{tabular}{lccccc}
        \toprule
        & \textbf{Ours} & \textbf{GAT} & \textbf{Transformer} & \textbf{Autoencoder} & \textbf{3D CNN} \\
        \midrule
        Max Memory Allocated (GB) & 37.70 & 29.00 & 9.30 & 6.20 & 4.30 \\
        Avg Time per Epoch (s)              & 1.63  & 0.84  & 1.69  & 0.58  & 0.56 \\
        Total Training Time (hours)        & 2.26  & 1.16  & 2.34  & 0.81  & 0.77 \\
        \bottomrule
    \end{tabular}
    \label{tab:E_compute}
    \end{small}
\end{table}

%% file: Conclusion.tex
\section{Conclusion}\label{sec:conc}

This study explores the use of a GraphGPS framework for the reconstruction of the 3D Riemann problems from sparse data, where 90\% of points are missing. The key feature of GraphGPS is that it combines the benefits of 1) message-passing modules, 2) modules that capture global context, and 3) positional encodings. For message-passing, we introduce a modified GAT layer, SA-GATConv, that leverages information on shocks and discontinuities in its attention mechanism. We also introduce guided message-passing, which sparsifies node aggregation by enforcing information to flow strictly from known nodes only. To make the model agnostic to the initialization of unknown nodes we make use of masked projection, which zeroes the representation of unknown nodes in the projected space of GraphGPS.

An ablation study on the modules of GraphGPS indicates SA-GATConv to be the superior message-passing module in terms of accuracy, enabling sharper reconstructions of shocks and discontinuities. From the ablation study we also find that the Mamba-2 module is the most efficient way for capturing global context whilst yielding acceptable accuracies. Experiments also show that guided-message passing reduces the time and memory required for reconstruction compared to (standard) dense message passing, whilst being on-par in terms of accuracy. Guided message-passing also exhibits more stable training than dense message-passing based on initial convergence behaviour. Furthermore we find that masked projection of the input yields similar accuracies to normal projection, even when the value of unknown nodes is pre-processed via feature propagation in the case of normal projection. Finally, we find that our GraphGPS framework outperforms a number of different ML benchmarks. 

The most notable limitation encountered in this study is the relatively large memory footprint of SA-GATConv for message-passing, despite its superior accuracy. This is due to the computation of the attention scores when aggregating information from neighbouring nodes. Future work should aim to alleviate this memory requirement such that finer resolutions could be explored. We also train a model for each 3D Riemann configuration separately, and it would be interesting to explore whether a model can generalize across multiple configurations, given a fixed distribution of known points. This study also assumes that known and unknown points/nodes are arranged as a uniform grid. In practice, grids tend to be non-uniform and this is therefore another direction to explore.

%% file: Appendix_A.tex
\section{HLLC Flux Estimation}\label{app:hllc}

\setcounter{figure}{0}

\begin{figure}[!ht]
\vskip 0.2in
\begin{center}
\centerline{\includegraphics[width=0.4\textwidth]{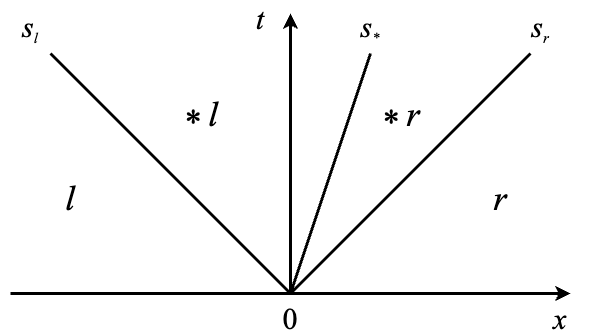}}
\caption{The three-wave model assumed by the HLLC solver. The left and right characteristic lines correspond to the fastest and slowest signals, $s_l$ and $s_r$, emerging from the cell interface at $x=0$. The middle characteristic corresponds to the wave of speed $s_*$ which accounts for contact and shear waves.}
\label{fig:hllc}
\end{center}
\vskip -0.2in
\end{figure}

An outline of the HLLC algorithm is presented in this section. In order to determine the HLLC flux normal to a cell interface for the 3D Euler equations, it is sufficient to consider a $x$-split version of the 3D Euler equations because of rotational invariance \cite{toro2019hllc}:
\begin{equation}
    {\partial_t}
\begin{pmatrix}
        \rho      \\
        \rho u      \\
        \rho v      \\
        \rho w \\
         E      \\
\end{pmatrix} + {\partial_x}
\begin{pmatrix}
        \rho u      \\
        \rho u^{2} + p \\
        \rho u v \\
        \rho u w \\
        u( E + p) \\
\end{pmatrix} = \mathbf{0} 
\end{equation}
We then estimate wave speeds $s_l$ and $s_r$ as \cite{toro2013riemann}:
\begin{equation}
    s_l = \Tilde{u} - \Tilde{a}, \,\,\,\,\,\,\,\, s_l = \Tilde{u} + \Tilde{a},
\end{equation}
where the Roe-average velocity and sound speed, $\Tilde{u}$ and $\Tilde{a}$, are defined as:
\begin{equation}
    \Tilde{u} = \frac{\sqrt{\rho_l}u_l + \sqrt{\rho_r}u_r}{\sqrt{\rho_l} + \sqrt{\rho_r}}, \,\,\,\,\,\,\,\, \Tilde{a} = \sqrt{\left(\gamma - 1\right)\left(\Tilde{H} - \frac{1}{2}\Tilde{u}^2\right)},
\end{equation}
and the Roe-average enthalpy $\Tilde{H}$ is defined as:
\begin{equation}
    \Tilde{H} =\frac{\sqrt{\rho_l}H_l + \sqrt{\rho_r}H_r}{\sqrt{\rho_l} + \sqrt{\rho_r}}, \,\,\,\,\,\,\,\, H =\frac{E + p}{\rho}.
\end{equation}
We then use $s_l$ and $s_r$ to compute intermediate speed $s_*$ \cite{toro2019hllc}:
\begin{equation}
    s_{*} = \frac{p_r - p_l + \rho_{l}u_{l}(s_l - u_l) 
    - \rho_{r}u_{r}(s_r - u_r)}{ \rho_{l}(s_l - u_l) 
    - \rho_{r}(s_r - u_r)}.
\end{equation}
The intermediate conservative vectors $\mathbf{U}_{*l}$ and $\mathbf{U}_{*r}$ are then computed using $s_*$, $s_l$ and $s_r$ as \cite{toro2019hllc}:
\begin{equation}
    \mathbf{U}_{*\eta} = \rho_{\eta}\left(\frac{s_\eta - u_\eta}{s_\eta - s_{*}}\right)\begin{pmatrix}
        1      \\
        s_{*}      \\
        v_\eta      \\
        w_\eta \\
        \frac{E_\eta}{\rho_\eta} + (s_{*} - u_\eta)\left[s_{*} + \frac{p_\eta}{\rho_\eta(s_\eta - u_\eta)}\right]      
\end{pmatrix},
\end{equation}
where $\eta = \left\{l,r\right\}$. Finally, the intermediate flux vectors $\mathbf{F}_{*l}$ and $\mathbf{F}_{*r}$ are computed from the state vectors so that the flux $\mathbf{F}_{i + \frac{1}{2}}^n$ is computed using equation \eqref{eq:HLLC}.
\begin{equation}\label{eq:HLLC}
    \mathbf{F}_{*\eta} = \mathbf{F}_\eta + s_\eta(\mathbf{U}_{*\eta} - \mathbf{U}_\eta), \;\;\;\;\;\;
    \mathbf{F}_{i + \frac{1}{2}}^t = \begin{cases}
        \mathbf{F}_l & 0 \leq s_l,\\
        \mathbf{F}_{*l} & s_l \leq 0 \leq s_{*},\\
        \mathbf{F}_{*r} &  s_{*} \leq 0 \leq s_{r},\\
        \mathbf{F}_{r} &   0 \geq s_{r}.
    \end{cases}
\end{equation}
Due to rotational invariance, an analogous procedure allows us to determine the $y$-flux $\mathbf{G}_{j + \frac{1}{2}}^t$ and the $z$-flux $\mathbf{H}_{k + \frac{1}{2}}^t$.

%% file: Appendix_B.tex
\section{WENO Reconstruction}\label{app:weno}

The weighted essentially non-oscillatory (WENO) scheme is a method of reconstructing cell averaged values to cell interfaces at a higher order of accuracy. In this section we briefly outline the 5th-order WENO scheme used in generating the ground truth $\mathbf{\bar{X}}$. We apply the WENO scheme on the conservative variables $\mathbf{U}$ for each spatial dimension separately. As such we demonstrate the reconstruction step for the $x$-dimension only. The procedure is analogous for the $y$- and $z$-dimensions.

The first step is to find candidate stencil approximations. For the 5th-order scheme these are:
\begin{align}
    &\bm{\alpha}_{0} = \frac{1}{3}\mathbf{U}_{i-2} -\frac{7}{6}\mathbf{U}_{i-1} - \frac{11}{6}\mathbf{U}_{i}\\[6pt]
    &\bm{\alpha}_{1} =-\frac{1}{6}\mathbf{U}_{i-1} + \frac{5}{6}\mathbf{U}_{i} + \frac{1}{3}\mathbf{U}_{i + 1} \\[6pt]
    &\bm{\alpha}_{2} = \frac{1}{3}\mathbf{U}_{i} + \frac{5}{6}\mathbf{U}_{i + 1} - \frac{1}{6}\mathbf{U}_{i + 2} 
\end{align}
We then obtain the smoothness indicators $\bm{\beta}$ as:
\begin{align}
    &\bm{\beta}_{0} = \frac{13}{12}\left(\mathbf{U}_{i-2}-2\mathbf{U}_{i-1}+\mathbf{U}_{i}\right)^{2} + \frac{1}{4}\left(\mathbf{U}_{i-2} - 4\mathbf{U}_{i-1} + 3\mathbf{U}_{i}\right)^{2} \\[6pt]
    &\bm{\beta}_{1} = \frac{13}{12}\left(\mathbf{U}_{i-1}-2\mathbf{U}_{i}+\mathbf{U}_{i+1}\right)^{2} + \frac{1}{4}\left(\mathbf{U}_{i-1} - \mathbf{U}_{i+1}\right)^{2}\\[6pt]
    &\bm{\beta}_{2} = \frac{13}{12}\left(\mathbf{U}_{i}-2\mathbf{U}_{i+1}+\mathbf{U}_{i+2}\right)^{2} + \frac{1}{4}\left(3\mathbf{U}_{i} - 4\mathbf{U}_{i+1} + \mathbf{U}_{i+2}\right)^{2}
\end{align}
We the find the non-linear weights $\bm{\omega}$:
\begin{equation}
   \bm{\omega}_n =  \frac{\bm{\zeta}_{n}}{\bm{\zeta}_{0} + \bm{\zeta}_{1} + \bm{\zeta}_{2}}, \,\,\,\,\,\,\,\, \bm{\zeta}_{n} = \frac{d_{n}}{\left(\epsilon + \bm{\beta}_{n}\right)^2}, \,\,\,\,\,\,\,\, \left(d_0, d_1, d_2\right) = \left(\frac{1}{10}, \frac{3}{5}, \frac{3}{10} \right), \,\,\,\,\,\,\,\, n \in \left\{0,1,2 \right\},
\end{equation}
where $\epsilon$ is a small number to prevent division by 0. The 5th-order approximation at the cell interface ${\mathbf{U}}_{i + \frac{1}{2}}^{\text{WENO}}$ can then be obtained:
\begin{equation}
    {\mathbf{U}}_{i + \frac{1}{2}}^{\text{WENO}} = \bm{\omega}_{0} \odot \bm{\alpha}_0  + \bm{\omega}_{1} \odot \bm{\alpha}_1  + \bm{\omega}_{2} \odot \bm{\alpha}_2 .
\end{equation}

%% file: Appendix_C.tex
\section{Initialization Values for 3D Riemann Problems}\label{app:init}

Initialization values are taken from \citet{balsara2015three} and \citet{hoppe2024systematic}.

\setcounter{table}{0}

\begin{table}[h]
\centering
\small
\caption{Octant definitions for domain $[0,1]^3$.}
\begin{tabular}{cccc}
\toprule
Octant & x & y & z \\
\midrule
1 & $>$ 0.5 & $>$ 0.5 & $>$ 0.5 \\
2 & $\leq$ 0.5 & $>$ 0.5 & $>$ 0.5 \\
3 & $\leq$ 0.5 & $\leq$ 0.5 & $>$ 0.5 \\
4 & $>$ 0.5 & $\leq$ 0.5 & $>$ 0.5 \\
5 & $>$ 0.5 & $>$ 0.5 & $\leq$ 0.5 \\
6 & $\leq$ 0.5 & $>$ 0.5 & $\leq$ 0.5 \\
7 & $\leq$ 0.5 & $\leq$ 0.5 & $\leq$ 0.5 \\
8 & $>$ 0.5 & $\leq$ 0.5 & $\leq$ 0.5 \\
\bottomrule
\end{tabular}
\label{tab:octants}

    \renewcommand{\arraystretch}{1.2}
    \caption{Case 1, ${J}_{21}^{}  {J}_{32}^{}  {S}_{34}^{-}  {S}_{41}^{+} {J}_{51}^{} {S}_{62}^{+} {J}_{73}^{} {S}_{84}^{-} {R}_{65}^{-} {S}_{76}^{-} {J}_{78}^{} {J}_{85}^{}$.}
    \begin{small}
    \begin{tabular}{cccccc}
        \toprule
        Octant & $\rho$ & $u$ & $v$ & $w$ & $p$ \\
        \midrule
        1 & 0.6223 & 0.1000 & -0.6259 & -0.1000 & 0.4000 \\
        2 & 0.5313 & 0.1000 & \,\,0.1000 & -0.8276 & 0.4000 \\
        3 & 0.6223 & 0.8259 & \,\,0.1000 & -0.1000 & 0.4000 \\
        4 & 1.2219 & 0.1000 & \,\,0.1000 & -0.1000 & 1.0683 \\
        5 & 0.5197 & 0.8259 & \,\,0.1000 & -0.1000 & 0.4000 \\
        6 & 1.0000 & 0.1000 & \,\,0.1000 & -0.1000 & 1.0000 \\
        7 & 0.5313 & 0.1000 & \,\,0.8276 & -0.1000 & 0.4000 \\
        8 & 0.6223 & 0.1000 & \,\,0.1000 & -0.6259 & 0.4000 \\
        \bottomrule
    \end{tabular}
    \end{small}

    \renewcommand{\arraystretch}{1.2}
    \caption{Case 2, ${S}_{21}^{+}  {S}_{32}^{-}  {S}_{34}^{-}  {S}_{41}^{+} {S}_{51}^{+} {S}_{62}^{-} {S}_{73}^{+} {S}_{84}^{-} {S}_{65}^{-} {S}_{76}^{+} {S}_{78}^{+} {S}_{85}^{-}$.}
    \begin{small}
    \begin{tabular}{cccccc}
        \toprule
        Octant & $\rho$ & $u$ & $v$ & $w$ & $p$ \\
        \midrule
        1 & 0.5065 & 0.0000 & 0.0000 & 0.0000 & 0.3500 \\
        2 & 1.1000 & 0.8939 & 0.0000 & 0.0000 & 1.1000 \\
        3 & 0.5065 & 0.8939 & 0.8939 & 0.0000 & 0.3500 \\
        4 & 1.1000 & 0.0000 & 0.8939 & 0.0000 & 1.1000 \\
        5 & 1.1000 & 0.0000 & 0.0000 & 0.8939 & 1.1000 \\
        6 & 0.5065 & 0.8939 & 0.0000 & 0.8939 & 0.3500 \\
        7 & 1.1000 & 0.8939 & 0.8939 & 0.8939 & 1.1000 \\
        8 & 0.5065 & 0.0000 & 0.8939 & 0.8939 & 0.3500 \\
        \bottomrule
    \end{tabular}
    \end{small}
\end{table}

\begin{table}[h]
    \centering
    \small
    \renewcommand{\arraystretch}{1.2}
    \caption{Case 3,${R}_{21}^{+}  {R}_{32}^{-}  {R}_{34}^{-}  {R}_{41}^{+} {S}_{51}^{-} {S}_{62}^{+} {S}_{73}^{-} {S}_{84}^{+} {R}_{65}^{-} {R}_{76}^{+} {R}_{78}^{+} {R}_{85}^{-}$.}
    \begin{small}
    \begin{tabular}{cccccc}
        \toprule
        Octant & $\rho$ & $u$ & $v$ & $w$ & $p$ \\
        \midrule
        1 & 1.0000 & \,\,0.0000 & \,\,0.0000 & -0.7881 & 1.0000 \\
        2 & 0.5000 & -0.7658 & \,\,0.0000 & -0.7881 & 0.3789 \\
        3 & 1.0000 & -0.7658 & -0.7658 & -0.7881 & 1.0000 \\
        4 & 0.5000 & \,\,0.0000 & -0.7658 & -0.7881 & 0.3789 \\
        5 & 0.5000 & \,\,0.0000 & \,\,0.0000 & \,\,0.0000 & 0.3789 \\
        6 & 1.0000 & -0.7658 & \,\,0.0000 & \,\,0.0000 & 1.0000 \\
        7 & 0.5000 & -0.7658 & -0.7658 & \,\,0.0000 & 0.3789 \\
        8 & 1.0000 & \,\,0.0000 & -0.7658 & \,\,0.0000 & 1.0000 \\
        \bottomrule
    \end{tabular}
    \end{small}

    \renewcommand{\arraystretch}{1.2}
    \caption{Case 4,${J}_{21}^{}  {J}_{32}^{}  {J}_{34}^{}  {J}_{41}^{} {S}_{51}^{-} {R}_{62}^{+} {S}_{73}^{-} {R}_{84}^{+} {J}_{65}^{} {J}_{76}^{} {J}_{78}^{} {J}_{85}^{}$.}
    \begin{small}
    \begin{tabular}{cccccc}
        \toprule
        Octant & $\rho$ & $u$ & $v$ & $w$ & $p$ \\
        \midrule
        1 & 1.5000 & \,\,0.1000 & -0.1500 & -0.2000 & 1.0000 \\
        2 & 1.0000 & \,\,0.1000 & \,\,0.1500 & -0.3000 & 1.0000 \\
        3 & 1.5000 & -0.1000 & \,\,0.1500 & -0.5000 & 1.0000 \\
        4 & 1.0000 & -0.1000 & -0.1500 & \,\,1.2000 & 1.0000 \\
        5 & 0.7969 & \,\,0.1000 & -0.1500 & \,\,0.3941 & 0.4000 \\
        6 & 0.5197 & \,\,0.1000 & \,\,0.1500 & -1.0259 & 0.4000 \\
        7 & 0.7969 & -0.1000 & \,\,0.1500 & \,\,0.0941 & 0.4000 \\
        8 & 0.5197 & -0.1000 & -0.1500 & \,\,0.4741 & 0.4000 \\
        \bottomrule
    \end{tabular}
    \end{small}

    \renewcommand{\arraystretch}{1.2}
    \caption{Case 5,${J}_{21}^{}  {J}_{32}^{}  {J}_{34}^{}  {J}_{41}^{} {J}_{51}^{} {J}_{62}^{} {J}_{73}^{} {J}_{84}^{} {J}_{65}^{} {J}_{76}^{} {J}_{78}^{} {J}_{85}^{}$.}
    \begin{small}
    \begin{tabular}{cccccc}
        \toprule
        Octant & $\rho$ & $u$ & $v$ & $w$ & $p$ \\
        \midrule
        1 & 0.5000 & -0.2500 & -0.5000 & -0.5000 & 1.0000 \\
        2 & 2.0000 & -0.2500 & \,\,0.5000 & -0.2500 & 1.0000 \\
        3 & 0.5000 & \,\,0.2500 & \,\,0.5000 & \,\,0.2500 & 1.0000 \\
        4 & 1.0000 & \,\,0.2500 & -0.5000 & -0.2500 & 1.0000 \\
        5 & 1.0000 & \,\,0.2500 & -0.2500 & -0.5000 & 1.0000 \\
        6 & 0.5000 & \,\,0.2500 & \,\,0.2500 & -0.2500 & 1.0000 \\
        7 & 2.0000 & -0.2500 & \,\,0.2500 & \,\,0.2500 & 1.0000 \\
        8 & 0.5000 & -0.2500 & -0.2500 & -0.2500 & 1.0000 \\
        \bottomrule
    \end{tabular}
    \end{small}

    \renewcommand{\arraystretch}{1.2}
    \caption{Case 6,${S}_{21}^{+}  {J}_{32}^{}  {J}_{34}^{}  {S}_{41}^{+} {S}_{51}^{+} {J}_{62}^{} {S}_{73}^{+} {J}_{84}^{} {J}_{65}^{} {S}_{76}^{+} {S}_{78}^{+} {J}_{85}^{}$.}
    \begin{small}
    \begin{tabular}{cccccc}
        \toprule
        Octant & $\rho$ & $u$ & $v$ & $w$ & $p$ \\
        \midrule
        1 & 0.5313 & \,\,0.0000 & \,\,0.0000 & \,\,0.0000 & 0.4000 \\
        2 & 1.0000 & \,\,0.7276 & \,\,0.0000 & \,\,0.0000 & 1.0000 \\
        3 & 0.8000 & \,\,0.0000 & \,\,0.0000 & -0.7276 & 1.0000 \\
        4 & 1.0000 & \,\,0.0000 & \,\,0.7276 & \,\,0.0000 & 1.0000 \\
        5 & 1.0000 & \,\,0.0000 & \,\,0.0000 & \,\,0.7276 & 1.0000 \\
        6 & 0.8000 & \,\,0.0000 & -0.7276 & \,\,0.0000 & 1.0000 \\
        7 & 1.0162 & -0.4014 & -0.4014 & -0.4014 & 1.4000 \\
        8 & 0.8000 & -0.7276 & \,\,0.0000 & \,\,0.0000 & 1.0000 \\
        \bottomrule
    \end{tabular}
    \end{small}
\end{table}

%% file: Appendix_D.tex
\section{Benchmark Models}\label{app:bm}

\setcounter{table}{0}

\begin{table}[h]
    \centering
    \small
    \caption{Architectural details of 3D CNN. For all Conv3D layers the kernel size is 3, the stride and padding are 1, and replicative padding is used.}
    \begin{tabular}{cccc}
        \toprule
        Layer No. & Layer Type & Input Channels & Output Channels \\
        \midrule
        1  & Conv3D + ReLU  & 5   & 16   \\
        2  & Conv3D + ReLU & 16  & 32   \\
        3  & Conv3D  + ReLU & 32  & 64  \\
        4-9  & Conv3D + ReLU & 64  & 64  \\
        10 & Conv3D + ReLU & 64  & 32  \\
        11 & Conv3D + ReLU & 32  & 16  \\
        12 & Conv3D  & 16  & 5  \\
        \bottomrule
    \end{tabular}
    \label{tab:CNN_BM}

    
    \caption{Architectural details of GAT. For all GATConv layers the number of attention heads is 1.}
    \begin{tabular}{cccc}
        \toprule
        Layer No. & Layer Type & Input Channels & Output Channels \\
        \midrule
        1  & GATConv + ReLU  & 10   & 64   \\
        2-9  & GATConv + ReLU & 64  & 64   \\
        10  & GATConv  & 64  & 10  \\
        \bottomrule
    \end{tabular}
    \label{tab:GNN_BM}

    
    \caption{Architectural details of autoencoder.}
    \begin{tabular}{cccc}
        \toprule
        Layer No. & Layer Type & Input Channels & Output Channels \\
        \midrule
        1  & Linear + ReLU  & 10   & 512   \\
        2  & Linear + ReLU & 512  & 256   \\
        3  & Linear + ReLU  & 256  & 128  \\
        4  & Linear + ReLU  & 128  & 64  \\
        5  & Linear + ReLU  & 64  & 32  \\
        6  & Linear + ReLU  & 32  & 64  \\
        7  & Linear + ReLU  & 64  & 128  \\
        8  & Linear + ReLU  & 128  & 256  \\
        9  & Linear + ReLU  & 256  & 512  \\
        10 & Linear & 512 & 10 \\
        \bottomrule
    \end{tabular}
    \label{tab:AE_BM}
     
     
    \caption{Architectural details of linear transformer. The number of heads for transformer layers is 16.}
    \begin{tabular}{cccc}
        \toprule
        Layer No. & Layer Type & Input Channels & Output Channels \\
        \midrule
        1  & Linear  & 10   & 64   \\
        2-6  & MLP + Transformer + MLP & 64  & 64   \\
        7  & Linear + ReLU  & 64  & 32  \\
        8  & Linear + ReLU  & 32  & 16  \\
        9  & Linear  & 16  & 10  \\
        \bottomrule
    \end{tabular}
    \label{tab:TRANS_BM}
\end{table}